\documentclass[10pt,journal,doublecolumn,twoside]{IEEEtran}
\IEEEoverridecommandlockouts

\usepackage{subfigure}
\usepackage{colortbl}
\usepackage{bm}

\usepackage{graphicx}  
\usepackage{url}       
\usepackage{rotating}
\usepackage{amsmath}   
\usepackage{cite}

\usepackage{amsfonts,amssymb}

\usepackage{stfloats}

\usepackage{cases}
\usepackage{algorithm}
\usepackage{multirow}
\usepackage{algorithmic}
\usepackage{amsthm}
\usepackage[utf8]{inputenc}

\newcommand{\myvec}[1]%
{\stackrel{\raisebox{-2pt}[0pt][0pt]
{\small$\rightharpoonup$}}{#1}}

\newcommand{\ls}[1]
    {\dimen0=\fontdimen6\the\font
     \lineskip=#1\dimen0
     \advance\lineskip.5\fontdimen5\the\font
     \advance\lineskip-\dimen0
     \lineskiplimit=.9\lineskip
     \baselineskip=\lineskip
     \advance\baselineskip\dimen0
     \normallineskip\lineskip
     \normallineskiplimit\lineskiplimit
     \normalbaselineskip\baselineskip
     \ignorespaces
    }

\newcounter{TempEqCnt}%

\usepackage{etoolbox}
\makeatletter
\patchcmd{\@makecaption}
  {\scshape}
  {}
  {}
  {}
\makeatother

\usepackage[absolute,showboxes]{textpos}
\TPMargin{3pt}

\begin{document}
\setlength{\columnsep}{0.24in}
\newcommand{\copyrightstatement}{
    \begin{textblock}{0.84}(0.08,0.95) 
         \noindent
         \footnotesize
         \copyright 2020 IEEE. Personal use of this material is permitted. Permission from IEEE must be obtained for all other uses, in any current or future media, including reprinting/republishing this material for advertising or promotional purposes, creating new collective works, for resale or redistribution to servers or lists, or reuse of any copyrighted component of this work in other works. DOI: 10.1109/TVT.2020.2966787.
    \end{textblock}
}
\copyrightstatement
\setlength{\columnsep}{0.24in}

\title{Joint OAM Multiplexing and OFDM in Sparse Multipath Environments
}

\author{Liping Liang,~\IEEEmembership{Student Member,~IEEE}, Wenchi Cheng,~\IEEEmembership{Senior Member,~IEEE},\\
Wei Zhang,~\IEEEmembership{Fellow,~IEEE}, and Hailin Zhang,~\IEEEmembership{Member,~IEEE}



\thanks{\ls{.5}Copyright (c) 2015 IEEE. Personal use of this material is permitted. However, permission to use this material for any other purposes must be obtained from the IEEE by sending a request to pubs-permissions@ieee.org.

A part of this work was presented in IEEE Global Communications Conference, 2018~\cite{2018_GC}. This work was supported in part by the National Natural Science Foundation of China under Grant 61771368, Young Elite Scientists Sponsorship Program By CAST under Grant 2016QNRC001, Doctoral Students' Short Term Study Abroad Scholarship Fund of Xidian University, and the Shenzhen Science \& Innovation Fund under grant JCYJ20180507182451820. (\emph{Corresponding author: Wenchi Cheng})

Liping Liang, Wenchi Cheng, and Hailin Zhang are with the State Key Laboratory of Integrated Services Networks, Xidian University, Xi'an, 710071, China (e-mails: lpliang@stu.xidian.edu.cn; wccheng@xidian.edu.cn; hlzhang@xidian.edu.cn).

Wei Zhang is with College of Electronics and Information Engineering, Shenzhen University, Shenzhen, China and School of Electrical Engineering and Telecommunications, the University of New South Wales, Sydney, Australia (email: weizhang@ieee.org).
}
}
\maketitle
\thispagestyle{empty}
\pagestyle{empty}

\begin{abstract}

The emerging orbital angular momentum (OAM) based wireless communications are expected to be a high spectrum-efficiency communication paradigm to solve the growing transmission data rate and limited bandwidth problem. Academic researchers mainly concentrate on the OAM-based line-of-sight (LoS) communications. However, there exist some surroundings around the transceiver in most practical wireless communication scenarios, thus forming multipath transmission.
In this paper, a hybrid orthogonal division multiplexing (HODM) scheme by using OAM multiplexing and orthogonal frequency division multiplexing (OFDM) in conjunction is proposed to achieve high-capacity wireless communications in sparse multipath environments, where the scatterers are sparse. We first build the OAM-based wireless channel in a LoS path and several reflection paths combined sparse multipath environments. We concentrate on less than or equal to three-time reflection paths because of the severe energy attenuation. The phase difference among the channel amplitude gains of the LoS and reflection paths, which is caused by the reflection paths, makes it difficult to decompose the OAM signals. We propose the phase difference compensation to handle this problem and then calculated the corresponding capacity in radio vortex wireless communications. Numerical results illustrate that the capacity of wireless communications by using our proposed HODM scheme can be drastically increased in sparse multipath environments.

\end{abstract}


\section{Introduction}

\IEEEPARstart{O}rbital angular momentum (OAM), different from the spin angular momentum of electromagnetic waves, is an interesting communication paradigm with high capacity and reliability in wireless communications~\cite{2010_Study,2018_OAM,2017_MFH,2017_generation,2015_imaging,2016_imaging,2018_detection}. When propagating along the same spatial axis, the beams with different integer OAM-modes, which are also referred to topological charges, are mutually orthogonal~\cite{2013_axis}. Thereby, multiple OAM-modes can be applied for several parallel data streams transmission without inter-mode interference theoretically. Hence, OAM is expected to be used for multiple users transmission where each user utilizes a different OAM-mode from others or for single user transmission where the user transmits data with OAM-modes multiplexing fashion.

\begin{figure*}
  \centering
  \includegraphics[width=0.9\textwidth]{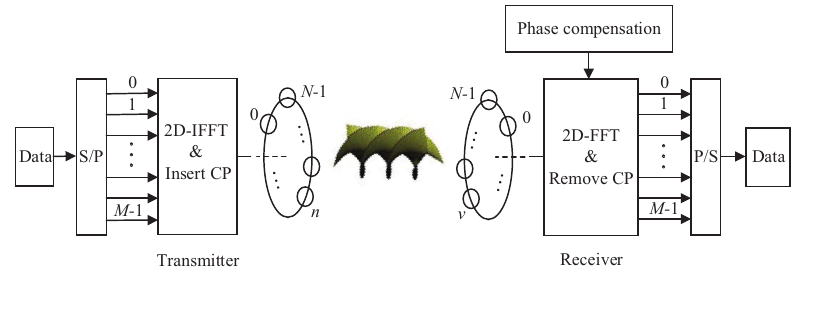}
  \caption{The system model in HODM communications. S/P: Serial-to-parallel converter; 2D-IFFT: Two-dimensional inverse fast Fourier transform; CP: Cyclic prefix; P/S: Parallel-to-serial converter.}
  \label{fig:sys}
  \vspace{-15pt}
\end{figure*}

OAM-mode multiplexing has been demonstrated to reach the goal of high capacity in a four OAM-modes multiplexing microwave communication experiment~\cite{2017_multiplexing}. Also, joint OAM-mode multiplexing and spatial multiplexing can achieve 16 Gbit/s line-of-sight (LoS) millimeter-wave communication at 1.8 meters transmission distance~\cite{2017_LOS}. Uniform circle array (UCA) based OAM is embedded into massive multiple input multiple output (MIMO) to achieve multiplicative capacity in both OAM-mode domain and spatial domain~\cite{2018_OEM}. In consequence, compatible with the frequency, code, spatial, and time, the emerging OAM-mode multiplexing provides the opportunities to significantly increase the capacity of wireless communications~\cite{2017_multiplexing,2017_LOS,2014_multiplexing,2016_polarization,2018_OEM}.

However, most existing researches mainly focus on studying OAM-mode multiplexing for LoS wireless communications~\cite{2017_short,2017_LOS}.
In some wireless communication scenarios, such as residential environments, there are a small number of randomly distributed surroundings around the transceivers, such as walls, doors, and so on. Thus, the signals are transmitted in LoS, reflection, diffuse, and scattering combined sparse multipath environments.
Since the power of signals after three-time reflections are relatively small~\cite{Goldsmith_2005}, we consider the sparse multipath environments with less than or equal to three-time reflection scenarios in this paper. Also, the comparison between OAM and MIMO is presented in~\cite{Sense}.

Academic researchers have shown much interest in orthogonal frequency division multiplexing (OFDM) technique, which is used for high capacity, but also for anti-multipath in wireless communications~\cite{2003_OFDM,2011_OFDM,2013_OFDM}. Utilizing the anti-multipath of OFDM, the inter-symbol interference can be absolutely canceled in radio vortex wireless communications. Thus, OFDM has been applied into a variety of scenarios in the past few decades, such as MIMO with spatial multiplexing~\cite{MIMO_IM} and full duplex with double spectrum efficiency~\cite{FD1,FD2,FD3}. Some existing academic researches have proved that OAM is compatible with the traditional OFDM to achieve extremely high capacity in wireless communications~\cite{2016_AFDM,2017_wCHENG,2014_AFDM}. The experiment has demonstrated that the signals carrying high order OAM-mode is susceptible to be affected by multipath interference~\cite{2016_AFDM}. By using the extra OAM-mode domain, the capacity can reach 230 bit/s/Hz in OFDM communications~\cite{2014_AFDM}. The authors proposed time-switched OFDM-OAM MIMO scheme to achieve high capacity of wireless communications while reducing computing complexity~\cite{OFDM_access}. 
However, these researches mainly focus on experimentally verifying the feasibility of joint OAM and OFDM to achieve high capacity of radio vortex wireless communications, which lacks the theoretical analysis of OAM signals transmission and decomposition. Also, they assume that the OAM-based wireless channel model in spares multipath environments is known and the inter-mode interference caused by reflection paths does not exist. We previously proposed joint OAM multiplexing and the traditional OFDM scheme, also referred to hybrid orthogonal division multiplexing (HODM), for anti-multipath while significantly increasing the capacity in multipath transmission, where we mainly focused on the signal transmission, detection, and capacity maximization of HODM communications~\cite{2018_GC}. Based on the previous work, this paper builds a specific OAM-based wireless channel model and mitigates the inter-mode interference in sparse multipath environments. Thereby, it is still challenging to use radio vortex wireless communications for sparse multipath transmission.

In this paper we propose a HODM scheme, thus not only enabling the efficient OAM-based transmission in sparse multipath environments, but also drastically increasing the capacity of wireless communications. We summarize the contributions as follows:

\begin{itemize}
  \item We build the OAM-based wireless channel model in sparse multipath environments containing a LoS path and several reflection paths. 

  \item We propose the phase difference compensation and theorematically analyze how to detect and decompose OAM signals in sparse multipath environments.

  \item We propose the HODM scheme to drastically increase the capacity of sparse multipath transmission in radio vortex wireless communications.
\end{itemize}

The remainder of this paper is organized as follows. Section~\ref{sec:sys} presents the HODM system model and builds the OAM-based wireless channel model. Section~\ref{sec:SE} proposes the HODM scheme, proposes the phase difference compensation to mitigate the inter-mode interference caused by multipath, and derives the maximum capacity of HODM communications. 
The performance of our proposed HODM scheme is evaluated in Section~\ref{sec:performance}. Conclusions are presented in Section~\ref{sec:conc}.
\section{System Model and Channel Model}\label{sec:sys}
In this section, we present the system model of our proposed HODM scheme and then build the OAM-based wireless channel model in sparse multipath environments containing a LoS path and several reflection paths.
\subsection{HODM System Model}

In this subsection, we investigate a HODM wireless communication as depicted in Fig.~\ref{fig:sys}, where a two-dimensional inverse fast Fourier transform (2D-IFFT) operator replaces the regular IFFT at the transmitter and a two-dimensional fast Fourier transform (2D-FFT) operator substitutes the traditional FFT at the receiver, respectively, in comparison with the traditional OFDM communications. Also, the HODM system model adds a phase difference compensation to the receiver. We consider $M$ parallel subcarriers and $N$ OAM-modes in HODM communications. As illustrated in Fig.~\ref{fig:sys}, the signals first are converted from serial state to parallel state and then processed by 2D-IFFT operator at the transmitter. A UCA with $N$ arrays equally spaced on the ring is applied to transmit the signals. To generate OAM-modes, the amplitudes of input signals for each array are same, but there is a consecutive phase difference from the first array to the last array for a given OAM-mode $l$ ($|l|\leq N/2$)~\cite{thide2007}. At the receiver, after removing cyclic prefix (CP) and compensating phase difference caused by multipath, 2D-FFT is used to decompose the signals. The carrier frequency modulated at the transmitter and demodulated at the receiver is assumed to be synchronized in this paper. 

\begin{figure}
  \centering
  \includegraphics[width=0.5\textwidth]{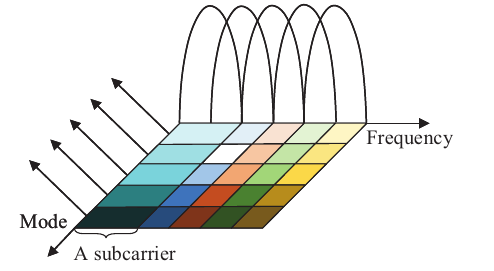}\\
  \caption{HODM transmission in both mode domain and frequency domain.}
  \label{fig:2D_mf}
  \vspace{-10pt}
\end{figure}

In wireless communications, OAM and OFDM are used for mode multiplexing and frequency multiplexing, respectively, to achieve high capacity. OAM and frequency are two independent domains. Therefore, we can jointly use OAM and OFDM, referred to HODM, in wireless communications. Thus, HODM modulated signals can be transmitted with $M$ subcarriers in the frequency domain and $N$ OAM-modes in the mode domain simultaneously as shown in Fig.~\ref{fig:2D_mf}, where each mode-frequency pair is identified by the specified color and neighbor subcarriers overlap each other. Hence, the high-rate data stream is divided into $MN$ parallel low-rate data streams for transmission in radio vortex wireless communications. Compared with the $M$ subchannels in conventional OFDM communications, HODM communications increase to $MN$ subchannels, thus significantly increasing the capacity of radio vortex wireless communications.

\begin{figure}
  \centering
  \includegraphics[width=0.52\textwidth]{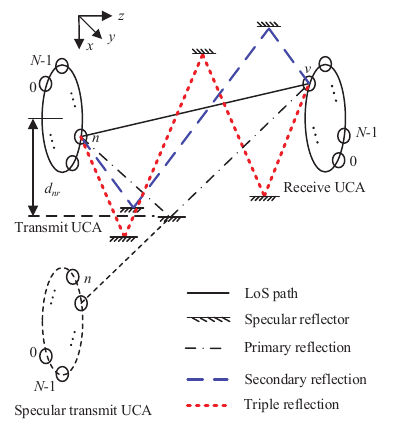}\\
  \caption{UCA-based OAM channel model consisting a LoS path and several reflection paths. }
  \label{fig:reflection}
  \vspace{-10pt}
\end{figure}

\subsection{UCA-Based OAM Channel Model}

In this paper, transmit signals are propagated in sparse multipath environments containing a LoS path, several reflection paths consisting of primary reflection paths, secondary reflection paths, and triple reflection paths. Figure~\ref{fig:reflection} shows the UCA-based OAM channel model, where the UCA based transmitter and receiver are coaxially aligned for simplicity. We assume that the reflection signals are caused by specular reflectors. Thus, the azimuthal angles of reflection signals can be considered to be the same with those of transmit signals. Referred to the traditional two-way channel model in wireless communications (\cite{Goldsmith_2005}, Chapter 2) and our previously derived LoS channel model \cite{JCIN_Jing}, the OAM-based wireless channel model in sparse multipath environments can be derived in the following.

We assume that there are $N_{nr}$ primary reflection paths, $N_{nt}$ secondary reflection paths, and $N_{ne}$ triple reflection paths. For each reflection, we can obtain the location of specular transmit UCA according to each specular reflector, which is parallel to the line between the centers of transmit and receive UCAs. Also, $D$, $r_{1}$, $r_{2}$, and $t$ are denoted by distance between the centers of transmit and receive UCAs, the radius of transmit UCA, the radius of receive UCA, and time variable, respectively.

For the LoS path, the expression of channel amplitude gain, denoted by $h_{vn,LoS,t}$, from the $n$-th $(0\leq n\leq N-1)$ transmit array to the $v$-th $(0\leq v\leq N-1)$ receive array is
\begin{eqnarray}
    h_{vn,LoS,t}=\frac{\beta \lambda}{4 \pi d_{vn} } e^{-j\frac{2 \pi d_{vn}}{\lambda}}\delta(t),
\end{eqnarray}
where $d_{vn}$ denotes the LoS distance between the $n$-th transmit array and the $v$-th receive array, $\beta$ represents the attenuation constant, $\lambda$ represents the carrier wavelength, and $\delta(t)$ represents the impulse response function with respect to $t$. The expression of $d_{vn}$ can be expressed as follows:
\begin{eqnarray}
    d_{vn}\!\!\!\!&=&\!\!\!\!\sqrt{D^{2}+r_{1}^{2}+r_{2}^{2}-2r_{1}r_{2}\cos\left(\frac{2 \pi v}{N}-\frac{2 \pi n}{N}\right)}\nonumber\\
    \!\!\!\!&=&\!\!\!\!\sqrt{D^{2}+r_{1}^{2}+r_{2}^{2}}\sqrt{1-\frac{2r_{1}r_{2}\cos\left(\frac{2 \pi v}{N}-\frac{2 \pi n}{N}\right)}{D^{2}+r_{1}^{2}+r_{2}^{2}}}.
    \label{eq:d_vn}
\end{eqnarray}
Based on Taylor series expansion and $r_{1}, r_{2} \ll D$, the second term on the right hand of Eq.~\eqref{eq:d_vn} is approximately expressed as follows:
\begin{equation}
    \sqrt{1-\frac{2r_{1}r_{2}\cos\left(\frac{2 \pi v}{N}-\frac{2 \pi n}{N}\right)}{D^{2}+r_{1}^{2}+r_{2}^{2}}} \approx 1-\frac{r_{1}r_{2}\cos\left(\frac{2 \pi v}{N}-\frac{2 \pi n}{N}\right)}{D^{2}+r_{1}^{2}+r_{2}^{2}}.
\end{equation}
Thus, using $\sqrt{D^{2}+r_{1}^{2}+r_{2}^{2}}$ for amplitude and $-r_{1}r_{2}\cos\left(\frac{2 \pi v}{N}-\frac{2 \pi n}{N}\right)\Big/\sqrt{D^{2}+r_{1}^{2}+r_{2}^{2}}$ for phase, we have 
\begin{equation}
    h_{vn,LoS,t}=\frac{\beta \lambda  e^{-j \frac{2 \pi \sqrt{D^{2}+r_{1}^{2}+r_{2}^{2}} }{\lambda }}}{4 \pi \sqrt{D^{2}+r_{1}^{2}+r_{2}^{2}}} e^{j \frac{2 \pi r_{1}r_{2}\cos(\frac{2 \pi v}{N}-\frac{2 \pi n}{N})}{\lambda  \sqrt{D^{2}+r_{1}^{2}+r_{2}^{2}}} }\delta(t).
    \label{eq:los}
\end{equation}

For the reflection paths, the reflection OAM signals carry the opposite modes after odd times reflections as compared with the initial LoS transmit OAM signals~\cite{2013experimental}.
Whereas, the reflection signals after even times reflections carry the same OAM-modes with the initial transmit OAM signals.

For primary reflection paths, we denote by $d_{nr}$ the distance between the transmit UCA center and the specular reflector for the $nr$-th $(nr=1,2,\cdots,N_{nr})$ primary reflection. Thus, we can derive the corresponding transmit distance $d_{vn,nr}$ between the $n$-th transmit array and the $v$-th receive array for the $nr$-th primary reflection as Eq.~\eqref{eq:d_vn_ref}.
\setcounter{TempEqCnt}{\value{equation}} 
\setcounter{equation}{4} 
\begin{figure*}[ht]
\begin{eqnarray}
 d_{vn,nr}&=&\sqrt{\left[2d_{nr}-r_{1}\cos\left(\frac{2 \pi n}{N}\right)-r_{2}\cos\left(\frac{2 \pi v}{N}\right)\right]^{2} + \left[r_{1}\sin\left(\frac{2 \pi n}{N}\right)-r_{2}\sin\left(\frac{2 \pi v}{N}\right)\right]^{2}+D^{2}} \nonumber\\
 &\approx & \sqrt{D^{2}+r_{1}^{2}+r_{2}^{2}+4d_{nr}^{2}}\left[1-\frac{-r_{1}r_{2}\cos\left(\frac{2 \pi}{N}v+\frac{2 \pi }{N}n\right)+2r_{2}d_{nr}\cos\left(\frac{2 \pi}{N}v\right)+2r_{1}d_{nr}\cos\left(\frac{2\pi}{N}n\right)}{D^{2}+r_{1}^{2}+r_{2}^{2}+4d_{nr}^{2}}\right].
 \label{eq:d_vn_ref}
\end{eqnarray}
\hrulefill
\end{figure*}
\begin{figure*}
\setcounter{equation}{8}
\begin{equation}
    h_{vn,nr,t}=\frac{R_{nr} \beta \lambda e^{-j\frac{2\pi \sqrt{D^{2}+r_{1}^{2}+r_{2}^{2}+4d_{nr}^{2}}}{\lambda}}}{4\pi \sqrt{D^{2}+r_{1}^{2}+r_{2}^{2}+4d_{nr}^{2}}}
    e^{j\frac{2\pi\left[-r_{1}r_{2}\cos\left(\frac{2\pi}{N}v+\frac{2\pi}{N}n\right)+2r_{1}d_{nr}\cos\left(\frac{2\pi}{N}n\right)+2r_{2}d_{nr}\cos\left
    (\frac{2\pi}{N}v\right)\right]}{\lambda \sqrt{D^{2}+r_{1}^{2}+r_{2}^{2}+4d_{nr}^{2}}}}\delta(t-\tau_{nr}),
    \label{eq:h_nr}
\end{equation}
\hrulefill
\end{figure*}
\begin{figure*}[ht]
\setcounter{equation}{10}
\begin{eqnarray}
 d_{vn,nt}\approx \sqrt{D^{2}+r_{1}^{2}+r_{2}^{2}+4d_{nt}^{2} }\left[1-\frac{r_{1}r_{2}\cos\left(\frac{2 \pi}{N}v-\frac{2 \pi }{N}n\right)-2r_{2}d_{nt}\cos\left(\frac{2 \pi}{N}v\right)+2r_{1}d_{nt}\cos\left(\frac{2\pi}{N}n\right)}{D^{2}+r_{1}^{2}+r_{2}^{2}+4d_{nt}^{2}}\right],
 \label{eq:d_vn_sec}
\end{eqnarray}
\hrulefill
\vspace{-10pt}
\end{figure*}

The reflection coefficient is one of important parameters of reflection channels. For different polarization transmissions, the expressions of the reflection coefficient are different. Our proposed HODM scheme can be for all kinds of polarization transmissions. Taking vertical polarization transmission as an example, the reflection coefficient, denoted by $R_{vn,nr}$, regarding the $n$-th array to the $v$-th receive array transmission for the specular reflector with a distance of $d_{nr}$ away from the transmit UCA center can be expressed as follows~\cite{Goldsmith_2005}:
\setcounter{equation}{5}
\begin{equation}
   R_{vn,nr}=\frac{\sin \alpha_{vn,nr} - \sqrt{\varepsilon_{nr}-\cos^{2}\alpha_{nr}/\varepsilon_{nr}}}{\sin \alpha_{vn,nr} + \sqrt{\varepsilon_{nr}-\cos^{2}\alpha_{vn,nr}/ \varepsilon_{nr}}},
\end{equation}
where $\alpha_{vn,nr}$ is the complementary angle of reflection angle for the $n$-th array to the $v$-th receive array transmission and $\varepsilon_{nr}$ represents the permittivity of the corresponding specular reflector, respectively, for the $nr$-th primary reflection path. Also, $ \alpha_{vn,nr}$ is given by
\begin{eqnarray}
     \alpha_{vn,nr} = \arcsin \frac{2d_{nr}-r_{1}\cos\left(\frac{2\pi}{N}n\right)-r_{2}\cos\left(\frac{2\pi }{N}v\right)}{d_{vn,nr}}.
\end{eqnarray}
For simplicity, the average reflection coefficient, denoted by $R_{nr}$, for the $nr$-th reflection path with respect to all transmit and receive arrays can be obtained as follows:
\setcounter{equation}{7}
\begin{eqnarray}
    R_{nr}=\frac{1}{N^2}\sum_{n=0}^{N-1}\sum_{v=0}^{N-1} R_{vn,nr}.
\end{eqnarray}
We denote by $c$ the speed of light. Due to $r_{1}, r_{2} < d_{nr} \ll D$, we can express the second term on the right hand of Eq.~\eqref{eq:d_vn_ref} as Taylor series. Then, $\sqrt{D^{2}+r_{1}^{2}+r_{2}^{2}+4d_{nr}^{2} }$ is used for amplitude. Thus, the channel amplitude gain, denoted by $h_{vn,nr,t}$, for the $n$-th array to the $v$-th receive array transmission corresponding to the specular reflector with a distance of $d_{nr}$ away from the transmit UCA center can be expressed as Eq.~\eqref{eq:h_nr},
where $\tau_{nr}$ is the time delay and given by
\setcounter{equation}{9}
\begin{eqnarray}
    \tau_{nr}=\frac{d_{vn,nr}-d_{vn}}{c}.
\end{eqnarray}

For the secondary reflection paths, we can obtain the first specular transmit UCA. Then, based on the first specular transmit UCA and the second reflector, the second specular transmit UCA can also be located. We denote by $d_{nt}^{(1)}$ and $d_{nt}^{(2)}$ the distances between the transmit UCA center and the first specular reflector as well as the second specular reflector for the $nt$-th ($nt=1,2,\cdots, N_{nt}$) reflection, respectively. The transmit distance, denoted by $d_{vn,nt}$, for the $n$-th array to the $v$-th receive array transmission can be calculated as Eq.~\eqref{eq:d_vn_sec}, where $d_{nt}$ is given by
\setcounter{equation}{11}
\begin{eqnarray}
d_{nt}=d_{nt}^{(1)}+d_{nt}^{(2)}.
\end{eqnarray}

To calculate the reflection coefficients of secondary reflection paths, we denote by $\varepsilon_{nt}^{(1)}$ and $\varepsilon_{nt}^{(2)}$ the permittivities of the first specular reflector and the second specular reflector, respectively. Since the specular reflectors are parallel to the line connected UCA centers of the transmitter and receiver, the reflection coefficient, denoted by $R_{vn,nt}$, for the $n$-th array to the $v$-th receive array transmission corresponding to the $nt$-th secondary reflection path can be approximately expressed as follows:
\begin{eqnarray}
    R_{vn,nt}=R_{vn,nt}^{(1)} R_{vn,nt}^{(2)},
\end{eqnarray}
where
\begin{eqnarray}
\begin{cases}
       R_{vn,nt}^{(1)}= \frac{\sin \alpha_{vn,nt} - \sqrt{\varepsilon_{nt}^{(1)}-\cos^{2}\alpha_{nt}/\varepsilon_{nt}^{(1)}}}{\sin \alpha_{vn,nt} + \sqrt{\varepsilon_{vn,nt}^{(1)}-\cos^{2}\alpha_{vn,nt}/ \varepsilon_{nt}^{(1)}}};
       \\
       R_{vn,nt}^{(2)}= \frac{\sin \alpha_{vn,nt} - \sqrt{\varepsilon_{nt}^{(2)}-\cos^{2}\alpha_{nt}/\varepsilon_{nt}^{(2)}}}{\sin \alpha_{vn,nt} + \sqrt{\varepsilon_{vn,nt}^{(2)}-\cos^{2}\alpha_{vn,nt}/ \varepsilon_{nt}^{(2)}}}.
\end{cases}
\label{eq:R_vn_nt}
\end{eqnarray}
In Eq.~\eqref{eq:R_vn_nt}, $\alpha_{vn,nt}$ is expressed as
\begin{eqnarray}
   \alpha_{vn,nt}=  \arcsin\frac{2d_{nt}-r_{1}\cos\left(\frac{2\pi}{N}n\right) - r_{2}\cos\left(\frac{2\pi }{N} v\right)}{d_{vn,nt}}.
\end{eqnarray}
For simplicity, we can obtain the average reflection coefficient, denoted by $R_{nt}$, for the $nt$-th reflection path with respect to all transmit and receive arrays as follows:
\begin{eqnarray}
    R_{nt}=\frac{1}{N^2}\sum_{n=0}^{N-1}\sum_{v=0}^{N-1} R_{vn,nt}.
\end{eqnarray}
\begin{figure*}
\setcounter{equation}{16}
\begin{eqnarray}
 h_{vn,nt,t}=\frac{R_{nt} \beta \lambda e^{-j\frac{2\pi \sqrt{D^{2}+r_{1}^{2}+r_{2}^{2}+4_{nt}^{2}}}{\lambda}}}{4\pi \sqrt{D^{2}+r_{1}^{2}+r_{2}^{2}+4d_{nt}^{2}}}
   e^{j\frac{2\pi\left[r_{1}r_{2}\cos\left(\frac{2\pi}{N}v-\frac{2\pi}{N}n\right)-2r_{1}d_{nt}\cos\left(\frac{2\pi}{N}n\right)+2r_{2}d_{nt}
    \cos\left(\frac{2\pi}{N }v\right)\right]}{\lambda \sqrt{D^{2}+r_{1}^{2}+r_{2}^{2}+4d_{nt}^{2}}}}\delta(t-\tau_{nt}),
\label{eq:h_nt}
\end{eqnarray}
\hrulefill
\end{figure*}
\begin{figure*}
\setcounter{equation}{22}
\begin{eqnarray}
  h_{vn,ne,t}&=&\frac{R_{ne} \beta \lambda e^{-j\frac{2\pi \sqrt{D^{2}+r_{1}^{2}+r_{2}^{2}+
  4d_{ne}^{2}}}{\lambda}}}{4\pi \sqrt{D^{2}+r_{1}^{2}+r_{2}^{2}+4d_{ne}^{2}}}
   e^{j\frac{2\pi\left[-r_{1}r_{2}\cos\left(\frac{2\pi}{N}v+\frac{2\pi}{N}n\right)+2r_{1}d_{ne}\cos\left(\frac{2\pi}{N}n\right)
    +2r_{2}d_{ne}
    \cos\left(\frac{2\pi}{N}v\right)\right]}{\lambda \sqrt{D^{2}+r_{1}^{2}+r_{2}^{2}+4d_{ne}^{2}}}} \delta(t-\tau_{ne}),
   \label{eq:h_ne}
\end{eqnarray}
\hrulefill
  \vspace{-10pt}
\end{figure*}
Then, the channel amplitude gain, denoted by $h_{vn,nt,t}$, for the $n$-th array to the $v$-th receive array transmission corresponding to $d_{nt}$ can be derived as Eq.~\eqref{eq:h_nt},
where $\tau_{nt}$ is the corresponding time delay and expressed as
\setcounter{equation}{17}
\begin{eqnarray}
    \tau_{nt}=\frac{d_{vn,nt}-d_{vn}}{c}.
\end{eqnarray}

For the triple reflection paths, we denote by $d_{ne}^{(1)}$, $d_{ne}^{(2)}$, and $d_{ne}^{(3)}$ the distances from the transmit UCA center to the first specular reflector, the second specular reflector, and the third specular reflector, respectively, for the $ne$-th $(ne=1,2,\cdots, N_{ne})$ triple reflection. Thus, the transmit distance, denoted by $d_{vn,ne}$, for the $n$-th array to the $v$-th receive array transmission for the $ne$-th triple reflection path can be calculated by replacing $d_{nr}$ by $d_{ne}$ in Eq.~\eqref{eq:d_vn_ref}, where
\setcounter{equation}{18}
\begin{equation}
   d_{ne}=d_{ne}^{(1)}+d_{ne}^{(2)}+d_{ne}^{(3)}.
\end{equation}
Also, $\varepsilon_{ne}^{(1)}$, $\varepsilon_{ne}^{(2)}$, and $\varepsilon_{ne}^{(3)}$ are denoted by the permittivities of the first specular reflector, the second specular reflector, and the third specular reflector, respectively. Thus, we can derive the corresponding average reflection coefficient, denoted by $R_{ne}$, corresponding to the $ne$-th triple reflection path as follows:
\setcounter{equation}{19}
\begin{eqnarray}
   R_{ne}&=&\frac{1}{N^2}\sum_{n=0}^{N-1}\sum_{v=0}^{N-1} R_{vn,ne}\nonumber
   \\
   &=& \frac{1}{N^2}\sum_{n=0}^{N-1}\sum_{v=0}^{N-1} R_{vn,ne}^{(1)}R_{vn,ne}^{(2)}R_{vn,ne}^{(3)},
\end{eqnarray}
where
\begin{eqnarray}
\setcounter{equation}{21}
\begin{cases}
       R_{vn,ne}^{(1)}= \frac{\sin \alpha_{vn,ne} - \sqrt{\varepsilon_{ne}^{(1)}-\cos^{2}\alpha_{ne}/\varepsilon_{ne}^{(1)}}}{\sin \alpha_{vn,ne} + \sqrt{\varepsilon_{vn,ne}^{(1)}-\cos^{2}\alpha_{vn,ne}/ \varepsilon_{ne}^{(1)}}};
       \\
       R_{vn,ne}^{(2)}= \frac{\sin \alpha_{vn,ne} - \sqrt{\varepsilon_{ne}^{(2)}-\cos^{2}\alpha_{ne}/\varepsilon_{ne}^{(2)}}}{\sin \alpha_{vn,ne} + \sqrt{\varepsilon_{vn,ne}^{(2)}-\cos^{2}\alpha_{vn,ne}/ \varepsilon_{ne}^{(2)}}};
       \\
      R_{vn,ne}^{(3)}= \frac{\sin \alpha_{vn,ne} - \sqrt{\varepsilon_{ne}^{(3)}-\cos^{2}\alpha_{ne}/\varepsilon_{ne}^{(3)}}}{\sin \alpha_{vn,ne} + \sqrt{\varepsilon_{vn,ne}^{(3)}-\cos^{2}\alpha_{vn,ne}/ \varepsilon_{ne}^{(3)}}}.
\end{cases}
\label{eq:R_vn_ne}
\end{eqnarray}
In Eq.~\eqref{eq:R_vn_ne}, $\alpha_{vn,ne}$ is derived as follows:
\begin{eqnarray}
   \alpha_{vn,ne}=  \arcsin\frac{2d_{ne}-r_{1}\cos\left(\frac{2\pi}{N}n\right) - r_{2}\cos\left(\frac{2\pi }{N} v\right)}{d_{vn,ne}}.
\end{eqnarray}
Then, the channel amplitude gain, denoted by $h_{vn,ne,t}$, for the $n$-th array to the $v$-th receive array transmission corresponding to $d_{ne}$ can be derived as Eq.~\eqref{eq:h_ne},
where
\setcounter{equation}{23}
\begin{eqnarray}
    \tau_{ne}=\frac{d_{vn,ne}-d_{vn}}{c}.
\end{eqnarray}

In Eqs. \eqref{eq:los}, \eqref{eq:h_nr}, \eqref{eq:h_nt}, and \eqref{eq:h_ne}, the channel amplitude gains of the paths are related to $v$, $n$, $\lambda$, and $t$, where $v$ and $n$ can be considered as the sample points of continuous signals in the angular domain. In OFDM communications, the channel model is converted from the time domain to the frequency domain after FFT at the receiver. In OAM communications, the received signals and channels are in the mode domain after OAM-based FFT. Therefore, the channel model is in the frequency and mode domains after 2D-FFT at the receiver, which can be directly dedicated to the HODM system.

\section{Interference Mitigation}~\label{sec:SE}
In this section, we develop the HODM scheme, mitigate the interference, and then calculate the maximum capacity of wireless communications. We first present the expression of the transmit signals in sparse multipath environments and insert the CP into transmit signals to mitigate the inter-symbol and inter-carrier interference. Next, we propose phase difference compensation to mitigate the inter-mode interference caused by multipath. Then, we demodulate the received signals in HODM communications. Finally, the conventional water-filling algorithm is applied to allocate the power, thus achieving the maximum capacity in wireless communications.


At the transmitter, the expression of the vortex signal, denoted by $x_{l,n}(t)$, with respect to the $l$-th OAM-mode for the $n$-th array is presented as follows:
 \begin{eqnarray}
  x_{l,n}(t)=\sum_{m=0}^{M-1}s_{l,m} e^{j 2 \pi f_{m} t}e^{j \frac{2 \pi n}{N} l},
 \end{eqnarray}
where $s_{l,m}$ is the transmit modulated signal for the $(l,m)$-th ($0 \leq m \leq M-1$) OAM-subcarrier block and $f_{m}$ is the $m$-th subcarrier frequency. Clearly, we can regard $x_{l,n}(t)$ as spatial sampling signal at the interval of $e^{j \frac{2 \pi n}{N} }$.

Thereby, we can model the continuous transmit signal, denoted by  $x_{l}(\phi,t)$, of $x_{l,n}(t)$ for the $(l,m)$-th OAM-subcarrier block as follows:
\begin{equation}
    x_{l}(\phi,t)=\sum_{m=0}^{M-1}s_{l,m} e^{j 2 \pi f_{m} t} e^{j \phi l},
\end{equation}
where we denote by $\phi \in [0, 2\pi]$ the azimuthal angle.


Then, we can obtain the emitted HODM-modulated signal, denoted by $x(\phi,t)$, corresponding to the whole OAM-modes within the transmit HODM duration as follows:
\begin{figure*}[ht]
\setcounter{equation}{32}
\begin{eqnarray}
Y_{v,k}&=&\sum_{n=0}^{N-1}h_{vn,LoS,k} X_{n,k}+\sum_{nr=1}^{N_{nr}}\sum_{n=0}^{N-1}h_{vn,nr,k} \widetilde{X}_{n-(L_{r}-nr),k-(L_{r}-nr)}^{'}
 +\sum_{nt=1}^{N_{nt}}\sum_{n=0}^{N-1}h_{vn,nt,k} X_{n-(L_{t}-nt),k-(L_{t}-nt)}\nonumber\\
&&+\sum_{ne=1}^{N_{ne}}\sum_{n=0}^{N-1}h_{vn,ne,k} \widetilde{X}_{n-(L_{e}-ne),k-(L_{e}-ne)}^{'}
+W_{v,k},
\label{eq:Y_v,k_rec}
\end{eqnarray}
\hrulefill
  \vspace{-10pt}
\end{figure*}
\begin{figure*}[ht]
\setcounter{equation}{34}
\begin{eqnarray}
    y_{l,m}\hspace{-0.2cm}&=&\hspace{-0.2cm}\frac{1}{MN}\sum_{k=0}^{M-1} \sum_{v=0}^{N-1} Y_{v,k} e^{-j \frac{2\pi vl}{N}}e^{-j\frac{2\pi mk}{M}}\nonumber\\
   \hspace{-0.2cm} &=&\hspace{-0.2cm} \!\!\frac{1}{MN}\!\!\sum_{k=0}^{M-1} \!\sum_{v=0}^{N-1}\!\sum_{n=0}^{N-1} \!\!h_{vn,LoS,k} X_{n,k}e^{-j \frac{2\pi vl}{N}}e^{-j\frac{2\pi mk}{M}}
   \!\! +\! \frac{1}{MN}\!\!\sum_{k=0}^{M-1}\! \sum_{v=0}^{N-1} \!\sum_{nr=1}^{N_{r}}\!\sum_{n=0}^{N-1}\!\! h_{vn,nr,k} \tilde{X}_{n\!-(L_{r}\!-nr),k\!-(L_{r}\!-nr)}^{'}e^{-\!j \frac{2\pi vl}{N}}e^{-\!j\frac{2\pi mk}{M}}\nonumber\\
   \hspace{-0.2cm} &+&\hspace{-0.2cm} \frac{1}{MN}\sum_{k=0}^{M-1}\sum_{v=0}^{N-1} \sum_{nt=1}^{N_{t}}\sum_{n=0}^{N-1} h_{vn,nt,k} X_{n-(L_{t}-nt),k-(L_{t}-nt)}e^{-j \frac{2\pi vl}{N}}e^{-j\frac{2\pi mk}{M}}\nonumber\\
   \hspace{-0.2cm} &+& \hspace{-0.2cm}\frac{1}{MN}\sum_{k=0}^{M-1} \sum_{v=0}^{N-1}\sum_{ne=1}^{N_{e}} \sum_{n=0}^{N-1}h_{vn,ne,k} \tilde{X}_{n-(L_{e}-ne),k-(L_{e}-ne)}^{'}e^{-j \frac{2\pi vl}{N}}e^{-j\frac{2\pi mk}{M}}+w_{l,m},
    \label{eq:y_ml_exp}
\end{eqnarray}
\hrulefill
  \vspace{-10pt}
\end{figure*}
\begin{eqnarray}
\setcounter{equation}{27}
  \hspace{-0.5cm}  x(\phi,t)\!\!\!\!&=&\!\!\!\!\!\sum_{l=\lfloor-\frac{N+2}{2}\rfloor}^{\lfloor\frac{N}{2}\rfloor} \!x_{l}(\phi,t)\nonumber\\
    \!\!\!&=&\!\! \!\!\sum_{m=0}^{M-1} \sum_{l=\lfloor-\frac{N+2}{2}\rfloor}^{\lfloor\frac{N}{2}\rfloor} s_{l,m}e^{j \phi l} e^{j 2\pi f_{m} t},
    \ 0 \leq t \leq T_{s},
    \label{eq:x}
    \end{eqnarray}
where $T_{s}$ denotes the transmit HODM duration and $\lfloor \cdot \rfloor$ represents the floor function.

We sample the emitted HODM-modulated signal $x(\phi,t)$, which means $t=\frac{kT_{s}}{M}$ ($0 \leq k \leq M-1$) in frequency domain and $\phi=\frac{2\pi n}{N}$ in spatial domain. Thus, we obtain
\setcounter{equation}{27}
\begin{eqnarray}
    X_{n,k}=\sum_{m=0}^{M-1} \sum_{l=\lfloor\frac{-N+2}{2}\rfloor}^{\lfloor\frac{N}{2}\rfloor} s_{l,m} e^{j \frac{2 \pi n l}{N} }e^{j \frac{2 \pi m k}{M}},
    \label{eq:X_kn}
\end{eqnarray}
where $X_{n,k}$ denotes the sampling signal in HODM communications. Clearly, the expression of the sampling HODM signal $X_{n,k}$ is the typical 2D-IFFT with respect to the emitted signal $s_{l,m}$.


Since there exists time delay of wireless channels caused by reflection paths, the CP is utilized to mitigate the interference. Thus, the expression of emitted signal, denoted by $\tilde{x}(\phi,t)$, with the insertion of CP is
\setcounter{equation}{28}
\begin{equation}
    \tilde{x}(\phi,t)=\!\!\sum_{m=0}^{M-1} \sum_{l=\lfloor\frac{-N+2}{2}\rfloor}^{\lfloor\frac{N}{2}\rfloor} s_{l,m}e^{j \phi l} e^{j 2\pi f_{m} t}, \ -T_{c} \leq t \leq T_{s},
\end{equation}
where the CP duration $T_{c}$ is larger than the maximum time delay, denoted by $\tau_{max}$, of wireless channels and $\tilde{x}(\phi,t)=x(\phi,t+T_{s})$ when $ -T_{c} \leq t \leq 0$. Thus, the corresponding sampling signal, denoted by $X_{n,u}$ $(u=-M_{c}, \cdots, M-1)$, can be expressed as follows:
\begin{eqnarray}
    X_{n,u}=\sum_{m=0}^{M-1} \sum_{l=\lfloor\frac{-N+2}{2}\rfloor}^{\lfloor\frac{N}{2}\rfloor} s_{l,m} e^{j \frac{2 \pi n l}{N} }}e^{j \frac{2 \pi u m}{M},
\end{eqnarray}
where $M_{c} \geq \lfloor\frac{M\tau_{max}}{T_{s}}\rfloor$ is the sampling length of CP. 

The expression of the received HODM signal, denoted by $y(\phi,t)$, for the whole receive arrays at the receiver is presented as follows:
\setcounter{equation}{30}
\begin{eqnarray}
    &&\hspace{-0.6cm}y(\phi,t)=h_{LoS}(\phi,t)\otimes\tilde{x}(\phi,t)+\sum_{nr=1}^{N_{nr}}h_{nr}(\phi,t)\otimes\tilde{x}^{'}(\phi,t)
    \nonumber \\
   &&\hspace{0.6cm}+\sum_{nt=1}^{N_{nt}}\!\! h_{nt}(\phi,t)\otimes\tilde{x}(\phi,t) \!+ \!\!\! \sum_{ne=1}^{N_{ne}}\! h_{ne}(\phi,t)\otimes\tilde{x}^{'}(\phi,t)
    \nonumber \\
   &&\hspace{0.6cm}+ W(\phi,t), \ \ \ \ \ \ \ \ \ \ \ \ \ \ \  -T_{c} \leq t \leq T_{s},
\end{eqnarray}
where $\otimes$ is the convolution operation, $W(\phi,t)$ denotes the received additive white Gaussian noise (AWGN) corresponding to the azimuthal angle $\phi$, and $h_{i}(\phi,t)\ (i=nr,nt,ne,LoS)$ denotes the channel response of the $i$-th path. Also, $\tilde{x}^{'}(\phi,t)$ represents the primary and triple reflection signal, which is given by
\begin{equation}
   \tilde{x}^{'}(\phi,t)=\sum\limits_{m=0}^{M-1} \sum_{l=\lfloor\frac{-N+2}{2}\rfloor}^{\lfloor\frac{N}{2}\rfloor} s_{l,m}e^{-j \phi l} e^{j 2\pi f_{m} t}, \ \ -T_{c} \leq t\leq T_{s}.
\end{equation}
Let $t=kT_{s}/M$ and $\phi={2\pi v}/{N}$ at the receiver. Thus, we have $W_{v,k}=W(\frac{2\pi v}{N},\frac{kT_{s}}{M})$. Next, the $M_{c}$ samples of CP are removed. The received sampling signal, denoted by $Y_{v,k}$, corresponding to $y(\phi,t)$ can be expressed as Eq.~\eqref{eq:Y_v,k_rec}, where
\setcounter{equation}{33}
\begin{eqnarray}
   \begin{cases}
    \!\!\widetilde{X}^{\prime}_{n-(L_{r}-nr),k-(L_{r}-nr)}\\
    \hspace{-0.1cm}=\!\!\!\sum\limits_{m=0}^{M-1}  \sum\limits_{l=\lfloor\!\frac{-N+2}{2}\!\rfloor}^{\lfloor\!\frac{N}{2}\!\rfloor}\!\!\! s_{l,m} e^{-j \frac{2 \pi [n\!-\!(L_{r}\!-\!nr)] l}{N} }\!e^{j \frac{2 \pi m[k\!-\!(L_{r}\!-\!nr)]}{M}};\\
   \!\! \widetilde{X}_{n-(L_{t}-nt),k-(L_{t}-nt)}\\
     \hspace{-0.1cm}=\!\!\sum\limits_{m=0}^{M-1} \sum\limits_{l=\lfloor\!\frac{-N+2}{2}\!\rfloor}^{\lfloor\!\frac{N}{2}\!\rfloor} \!\!\!s_{l,m} e^{j \frac{2 \pi [n\!-\!(L_{t}\!-\!nt)] l}{N} }e^{j \frac{2 \pi m[k-(L_{t}-nt)]}{M}};\\
    \!\!\widetilde{X}^{\prime}_{n-(L_{e}-ne),k\!-\!(L_{e}\!-\!ne)}\\
     \hspace{-0.1cm}=\!\!\sum\limits_{m=0}^{M-1} \sum\limits_{l=\lfloor\!\frac{-N+2}{2}\!\rfloor}^{\lfloor\!\frac{N}{2}\!\rfloor} \!\!\!s_{l,m} e^{-j \frac{2 \pi [n\!-\!(L_{e}\!-\!ne)] l}{N} }e^{j \frac{2 \pi m[k\!-\!(L_{e}\!-\!ne)]}{M}},
   \end{cases}
\end{eqnarray}
$L_{r}$, $L_{t}$, and $L_{e}$ are the maximum normalized values of $\tau_{nr}$, $\tau_{nt}$, and $\tau_{ne}$, respectively, by the sample interval $T_{s}/M$.

\begin{figure*}
\setcounter{equation}{40}
\begin{eqnarray}
\begin{aligned}
    &\frac{1}{MN}\sum_{k=0}^{M-1} \sum_{v=0}^{N-1}\sum_{nr=1}^{N_{r}}\sum_{n=0}^{N-1} h_{vn,nr,k}\widetilde{X}_{n-(L_{r}-nr),k-(L_{r}-nr)}^{'} e^{-j \frac{2\pi vl}{N}}e^{-j\frac{2\pi mk}{M}}
\\
    & \overset{N \rightarrow \infty}{\approx} \!\sum_{m^{'}=0}^{M-1}\!\sum_{l^{\prime}=\lfloor\frac{-N+2}{2}\rfloor}^{\lfloor\frac{N}{2}\rfloor} \!\! \!  \frac{s_{l^{'},m^{'}}}{MN} \!\!\!\sum_{k=0}^{M-1} \! \sum_{v=0}^{N-1} \sum_{nr=1}^{N_{r}}  \! \!\frac{ R_{nr}\beta \lambda e^{-j\frac{2\pi \sqrt{D^{2}+r_{1}^{2}+r_{2}^{2}+4d_{nr}^{2}}}{\lambda}}}{4\pi\sqrt{D^{2}+r_{1}^{2}+r_{2}^{2}+4d_{nr}^{2}}} \! \frac{N}{2\pi}\!\! \int_{0}^{2\pi} \! \! e^{j\frac{2\pi r_{1}\sqrt{r_{2}^{2}+4d_{nr}^{2}-4r_{2}h_{nr}\cos\left(\frac{2\pi v}{N}\right)}\cos{\phi^{'}}}{\lambda \!\sqrt{D^{2}+r_{1}^{2}+r_{2}^{2}+4d_{nr}^{2}}}} \!e^{-j\phi^{'}\! l^{'}}\! \!d \phi^{'}
\\
    & \hspace{3.2cm} e^{-j\arctan{\frac{r_{2} \sin\left(\frac{2\pi v}{N}\right)}{2d_{nr}-r_{2} \cos\left(\frac{2\pi v}{N}\right)}}l^{\prime}} e^{j\frac{4\pi r_{2} d_{nr} \cos\left(\frac{2\pi v}{N}\right) }{\lambda \sqrt{D^{2}+r_{1}^{2}+r_{2}^{2}+4d_{nr}^{2}}}} e^{j\frac{2\pi (L_{r}-nr)l^{'}}{N}} e^{-j\frac{2\pi (L_{r}-nr)m^{'}}{M}} e^{-j\frac{2\pi vl }{N}}  e^{j\frac{2\pi k (m^{'}-m)}{M}}
\\
    & = \sum_{m^{'}=0}^{M-1}\sum_{l^{\prime}=\lfloor\frac{-N+2}{2}\rfloor}^{\lfloor\frac{N}{2}\rfloor} \! \! \!  \frac{s_{l^{'},m^{'}}}{MN} \sum_{k=0}^{M-1} \sum_{v=0}^{N-1} \sum_{nr=1}^{N_{r}} \!\frac{ R_{nr}\beta \lambda e^{-j\frac{2\pi \sqrt{D^{2}+r_{1}^{2}+r_{2}^{2}+4d_{nr}^{2}}}{\lambda}}}{4\pi\sqrt{D^{2}+r_{1}^{2}+r_{2}^{2}+4d_{nr}^{2}}}
    N j^{-l^{'}} J_{l^{'}}\left(\frac{2\pi r_{1}\sqrt{r_{2}^{2}+4d_{nr}^{2}-4r_{2}d_{nr}\cos\left(\frac{2\pi v}{N}\right)}}{\lambda \sqrt{D^{2}+r_{1}^{2}+r_{2}^{2}+4d_{nr}^{2}}}\right)
\\
    & \hspace{3.2cm} e^{-j\arctan{\frac{r_{2} \sin\left(\frac{2\pi v}{N}\right)}{2d_{nr}-r_{2} \cos\left(\frac{2\pi v}{N}\right)}}l^{\prime}} e^{j\frac{4\pi r_{2} d_{nr} \cos\left(\frac{2\pi v}{N}\right) }{\lambda \sqrt{D^{2}+r_{1}^{2}+r_{2}^{2}+4d_{nr}^{2}}}} e^{j\frac{2\pi (L_{r}-nr)l^{'}}{N}} e^{-j\frac{2\pi (L_{r}-nr)m^{'}}{M}} e^{-j\frac{2\pi vl }{N}}  e^{j\frac{2\pi k (m^{'}-m)}{M}}.
\end{aligned}
\label{eq:y_ml_pri_error}
\end{eqnarray}
\hrulefill
  \vspace{-10pt}
\end{figure*}

We denote by $y_{l,m}$ the demodulated signal with respect to the $(l,m)$-th OAM-subcarrier block. Then, $y_{l,m}$ is calculated as Eq.~\eqref{eq:y_ml_exp}, where $w_{l,m}$ is given by
\setcounter{equation}{35}
\begin{eqnarray}
    w_{l,m}=\frac{1}{MN}\sum_{k=0}^{M-1}\sum_{v=0}^{N-1}W_{v,k}e^{-j \frac{2\pi vl}{N}}e^{-j\frac{2\pi mk}{M}}.
    \label{eq:w_l,m}
\end{eqnarray}
 Thereby, we have the following Theorem 1.

\emph{\textbf{Theorem 1}}: The channel amplitude gain, denoted by $h_{l,m,LoS}$, of the LoS path for the $(l,m)$-th OAM-subcarrier block is
\begin{equation}
    h_{l,m,LoS}\!\!=\!\!\frac{\beta \lambda_{m} N j^{-l} e^{-j\frac{2\pi \sqrt{D^{2}+r_{1}^{2}+r_{2}^{2}}}{\lambda_{m}}}}{\sqrt{D^{2}+r_{1}^{2}+r_{2}^{2}}} \!\! J_{l}\!\!\left(\!\!\frac{2\pi r_{1} r_{2}}{\lambda_{m} \! \sqrt{D^{2}\!+\!r_{1}^{2}\!+\!r_{2}^{2}}}\!\!\right),
\end{equation}
where
\setcounter{equation}{37}
\begin{eqnarray}
    J_{l}(z)=\frac{j}{2\pi } \int_{0}^{2\pi} e^{j(z\cos \psi-l\psi)} d\psi
\end{eqnarray}
is the $l$ order Bessel function corresponding to the first kind and $\lambda_{m}$ is the $m$-th carrier wavelength.
\begin{proof}
See Appendix~\ref{Appendix:_A}.
\end{proof}
Observing Eq.~\eqref{eq:h_nr}, we have
\begin{equation}
   \begin{aligned}
    &-r_{1}r_{2}\cos\left(\frac{2\pi}{N}v+\frac{2\pi}{N}n\right)+2r_{1}d_{nr}\cos\left(\frac{2\pi}{N}n\right)
    \\
    & = r_{1} \sqrt{r_{2}^{2}+4d_{nr}^{2}-4r_{2}d_{nr}^{2}\cos \left(\frac{2\pi}{N}v\right)} \cos\left(\frac{2\pi}{N}n-\theta\right),
   \end{aligned}
\end{equation}
where
\begin{equation}
 \theta = \arctan \frac{r_{2} \sin\left(\frac{2\pi v}{N}\right)}{2d_{nr}-r_{2}\cos\left(\frac{2\pi v}{N}\right)}.
\end{equation}
Then, the second term on the right hand of Eq.~\eqref{eq:y_ml_exp} corresponding to the primary reflections can be calculated as Eq.~\eqref{eq:y_ml_pri_error}. Clearly, when $m^{'}=m$, the corresponding $m$-th subcarrier signal can be obtained. However, the corresponding $l$-th OAM-mode signal cannot be obtained due to $e^{-j\arctan{\left[r_{2} \sin\left(\frac{2\pi v}{N}\right)\right]/ \left[2d_{nr}-r_{2} \cos\left(\frac{2\pi v}{N}\right)\right]}l^{\prime}}$ and $e^{j\left[4\pi r_{2} d_{nr} \cos\left(\frac{2\pi v}{N}\right)\right] / \left(\lambda \sqrt{D^{2}+r_{1}^{2}+r_{2}^{2}+4d_{nr}^{2}}\right)} $. The corresponding OAM signals for the secondary reflections and triple reflections also cannot be obtained.

To decompose the OAM signal successfully, a compensation factor for reflection channel amplitude gains is needed. Comparing the expressions among $h_{vn,LoS,t}$, $h_{vn,nr,t}$, $h_{vn,nt,t}$, and $h_{vn,ne,t}$, we can find that the last three terms add an exponential factor based on the first term, respectively.
Therefore, we set an exponential factor, denoted by $h_{vn,e}$, for the $n$-th array to the $v$-th receive array transmission to compensate the phase difference as follows:
\setcounter{equation}{41}
\begin{eqnarray}
    h_{vn, e}= e^{j\frac{2\pi \left[2(-1)^{\mu}r_{1}d_{max}\cos\left(\frac{2\pi n}{N}\right)-2r_{2}d_{max}\cos\left(\frac{2\pi v}{N}\right)\right]}{\lambda \sqrt{D^{2}+r_{1}^{2}+r_{2}^{2}+4d_{max}^{2}}}},
\end{eqnarray}
where $d_{max}$ is the maximum sum of distances from each specular transmit UCA center to the transmit UCA center among the all reflection paths and $\mu$ is one for primary as well as triple reflections and zero for secondary reflections. The exponential factor can be obtained by using ray tracing method.
Hence, we can re-express $Y_{v,k}$ as follows:
\begin{eqnarray}
   Y_{v,k}\hspace{-0.3cm}&=& \hspace{-0.3cm}\sum_{n=0}^{N-1}h_{vn,LoS,k} X_{n,k}\nonumber\\
   \hspace{-0.3cm}&+&\hspace{-0.3cm}\sum_{nr=1}^{N_{nr}}\sum_{n=0}^{N-1}h_{vn,nr,k} h_{vn,e} \widetilde{X}_{n-(L_{r}-nr),k-(L_{r}-nr)}^{'} \nonumber\\
\hspace{-0.3cm}&+&\hspace{-0.3cm} \sum_{nt=1}^{N_{nt}}\sum_{n=0}^{N-1}h_{vn,nt,k}  h_{vn,e}X_{n-(L_{t}-nt),k-(L_{t}-nt)}\nonumber\\
\hspace{-0.3cm}&+&\hspace{-0.3cm}\!\!\sum_{ne=1}^{N_{ne}}\sum_{n=0}^{N-1}h_{vn,ne,k}  h_{vn,e} \widetilde{X}_{n-(L_{e}-ne),k-(L_{e}-ne)}^{'}
\!+\!W_{v,k}.
\nonumber\\
\label{eq:y_vk_com}
\end{eqnarray}
With the compensation, the decomposed signal of the primary reflections, secondary reflections, and triple reflections for each OAM-subcarrier block can be derived, respectively. We denote by $h_{l,m,N_{r}}$, $h_{l,m,N_{t}}$, and $h_{l,m,N_{e}}$ the channel amplitude gains of $N_{r}$ primary reflections, $N_{t}$ secondary reflections, and $N_{e}$ triple reflections, respectively, for the $(l,m)$-th OAM-subcarrier block. Then, the Theorem 2 is obtained as follows.
\begin{figure*}
\setcounter{equation}{43}
\begin{eqnarray}
 h_{l,m,r}&=& h_{l,m,N_{r}}+h_{l,m,N_{t}}+h_{l,m,N_{e}}
 \nonumber\\
&=& \sum\limits_{nr=1}^{N_{r}} \frac{ R_{nr}\beta \lambda_{m} N j^{-3l} e^{-j\frac{2\pi \sqrt{D^{2}+r_{1}^{2}+r_{2}^{2}+4d_{nr}^{2}}}{\lambda_{m}}}}{\sqrt{D^{2}+r_{1}^{2}+r_{2}^{2}+4d_{nr}^{2}}}  J_{l}\left(\frac{2\pi r_{1} r_{2}}{\lambda_{m} \sqrt{D^{2}+r_{1}^{2}+r_{2}^{2}+4d_{nr}^{2}}}\right) e^{j\frac{2\pi l (L_{r}-nr)}{N}} e^{-j\frac{2\pi m (L_{r}-nr)}{M}}
\nonumber\\
 &+& \sum\limits_{nt=1}^{N_{t}}\frac{ R_{nt}\beta \lambda_{m} N j^{-l} e^{-j\frac{2\pi \sqrt{D^{2}+r_{1}^{2}+r_{2}^{2}+4d_{nt}^{2}}}{\lambda_{m}}}}{\sqrt{D^{2}+r_{1}^{2}+r_{2}^{2}+4d_{nt}^{2}}}  J_{l}\left(\frac{2\pi r_{1} r_{2}}{\lambda_{m} \sqrt{D^{2}+r_{1}^{2}+r_{2}^{2}+4d_{nt}^{2}}}\right) e^{-j\frac{2\pi l (L_{t}-nt)}{N}} e^{-j\frac{2\pi m (L_{t}-nt)}{M}}
\nonumber\\
 &+&\sum\limits_{ne=1}^{N_{e}} \frac{ R_{ne}\beta \lambda_{m} N j^{-3l} e^{-j\frac{2\pi \sqrt{D^{2}+r_{1}^{2}+r_{2}^{2}+4d_{ne}^{2}}}{\lambda_{m}}}}{\sqrt{D^{2}+r_{1}^{2}+r_{2}^{2}+4d_{ne}^{2}}}  J_{l}\left(\frac{2\pi r_{1} r_{2}}{\lambda_{m} \sqrt{D^{2}+r_{1}^{2}+r_{2}^{2}+4d_{ne}^{2}}}\right) e^{j\frac{2\pi l (L_{e}-ne)}{N}} e^{-j\frac{2\pi m (L_{e}-ne)}{M}}.
\label{eq:h_ml_ref}
\end{eqnarray}
\hrulefill
\end{figure*}
\begin{figure*}
\setcounter{equation}{50}
             \begin{equation}
        \Delta_{l} =\frac{\rho^{2}\bm{H}_{l}^{H}\bm{H}_{l}\mathbb{E}[\bm{s}_{l}\bm{s}_{l}^{H}]\mathbb{E}\left[\bm{\Omega}_{l} \bm{s}_{l}\bm{s}_{l}^{H}\bm{\Omega}_{l}^{H}
        + \bm{\Omega}_{l}\bm{H}_{l}^{-1}\bm{w}_{l}\bm{w}_{l}^{H}(\bm{H}_{l}^{-1})^{H}\bm{\Omega}_{l}^{H}
        \right]}{\mathbb{E}[\bm{w}_{l}\bm{w}_{l}^{H}]\mathbb{E}\left[\bm{w}_{l}\bm{w}_{l}^{H}
        +\rho^{2} \bm{\Omega}_{l} \bm{s}_{l}\bm{s}_{l}^{H}\bm{\Omega}_{l}^{H}
        +\ \rho^{2}\bm{\Omega}_{l}\bm{H}_{l}^{-1}\bm{w}_{l}\bm{w}_{l}^{H}(\bm{H}_{l}^{-1})^{H}\bm{\Omega}_{l}^{H}
        \right]}.
        \label{eq:loss}
     \end{equation}
     \hrulefill
       \vspace{-10pt}
\end{figure*}

\emph{\textbf{Theorem 2}}: For a given number of reflection paths, the channel amplitude gain of reflection paths, denoted by $h_{l,m,r}$, with respect to the $(l,m)$-th OAM-subcarrier block is presented as Eq.~\eqref{eq:h_ml_ref}.
\begin{proof}
    See Appendix~\ref{Appendix:_C}.
\end{proof}

Therefore, $y_{l,m}$ is re-expressed as follows:
\setcounter{equation}{44}
\begin{eqnarray}
    y_{l,m}&=& (h_{l,m,LoS}+h_{l,m,r}) s_{l,m} + w_{l,m} \nonumber
    \\
    &=& h_{l,m} s_{l,m} + w_{l,m},
\label{eq:y_lm_error}
\end{eqnarray}
where $h_{l,m}$ is the channel amplitude gain for the $(l,m)$-th OAM-subcarrier block in sparse multipath environments.



In the following, $(\cdot)^{T}$ and $(\cdot)^{-1}$ represent the transpose and inverse of a matrix, respectively. For the $l$-th OAM-mode, we denote by ${\bf y}_{l}=(y_{l,0},y_{l,1},\cdots,y_{l,M-1})^{T}$ and ${\bf w}_{l}=(w_{l,0},w_{l,1},\cdots,w_{l,M-1})^{T}$ the received signal vector and noise vector, respectively, with respect to the $l$-th OAM-mode. Assuming that $\rho\bm{\Omega}_{l}$ is the channel estimation error (CEE) for the $l$-th OAM-mode, where $\rho \ll 1$ represents the accuracy of channel estimation and $\bm{\Omega}_{l}$ follows i.i.d. zero-mean Gaussian distribution \cite{CEE,imperfect}, we have 
     \begin{eqnarray}
         \hat{\bm{s}}_{l}=(\bm{H}_{l}+\rho\bm{\Omega}_{l})^{-1}\bm{y}_{l},
     \end{eqnarray}
     where $\hat{\bm s}_{l}$ and ${\bm s}_{l}=(s_{l,0},s_{l,1},\cdots,s_{l,M-1})^{T}$ denote the transmit signal estimation vector and the transmit signal vector, respectively, for the $l$-th OAM-mode. In addition, ${\bf H}_{l}={\rm{diag}}\{h_{l,0},h_{l,1},\cdots,h_{l,M-1}\}$. $\rho=0$ implies the perfect channel estimation.

     Using the linear part of Taylor series (\cite{CEE}, Eq. (9)), we have
    \begin{eqnarray}
       \hspace{-0.3cm} \hat{\bm{s}}_{l}\hspace{-0.25cm}&=&\hspace{-0.25cm}\bm{H}_{l}^{-1}(\bm{I}_{M}-\rho\bm{\Omega}_{l}\bm{H}_{l})\bm{y}_{l}
        \nonumber\\
        \hspace{-0.25cm}&=&\hspace{-0.25cm}\bm{s}_{l}+ \bm{H}_{l}^{-1}\bm{w}_{l}-\rho\bm{H}_{l}^{-1}\bm{\Omega}_{l}\bm{s}_{l}-\rho\bm{H}_{l}^{-1}\bm{\Omega}_{l}\bm{H}_{l}^{-1}\bm{w}_{l}.
        \label{eq:noise}
    \end{eqnarray}

     The last two terms on the right hand of Eq.~\eqref{eq:noise} are the additional interference and noise under imperfect channel estimation. Thus, we have the noise, denoted by $\widehat{\bm{w}}_{l}$, after zero-forcing detection as follows:
     \begin{eqnarray}
       \widehat{\bm{w}}_{l}=\bm{H}_{l}^{-1}\bm{w}_{l}-\rho\bm{H}_{l}^{-1}\bm{\Omega}_{l}\bm{s}_{l}-\rho\bm{H}_{l}^{-1}\bm{\Omega}_{l}\bm{H}_{l}^{-1}\bm{w}_{l}.
     \end{eqnarray}
     The covariance of $\widehat{\bm{w}}_{l}$, is derived as follows:
     \begin{eqnarray}
        \mathbb{E}[\bm{w}_{l}\bm{w}_{l}^{H}]=\mathbb{E}\left[\bm{w}_{l}\bm{w}_{l}^{H}
        + \rho^{2}\bm{\Omega}_{l}\bm{H}_{l}^{-1}\widehat{\bm{w}}_{l}\widehat{\bm{w}}_{l}^{H}(\bm{H}_{l}^{-1})^{H}\bm{\Omega}_{l}^{H}\right.
        \nonumber\\
            \left. +\rho^{2} \bm{\Omega}_{l} \bm{s}_{l}\bm{s}_{l}^{H}\bm{\Omega}_{l}^{H}
        \right](\bm{H}_{l}^{H}\bm{H}_{l})^{-1},
        \label{eq:vari}
     \end{eqnarray}
     where $(\cdot)^{H}$ represents the conjugate transpose of a matrix. Observing Eq.~\eqref{eq:vari}, we can find that the covariance of the noise in the present of CEE is larger than that under perfect channel estimation.

     The received signal-to-noise ratio (SNR), denoted by $\gamma_{l}$, with respect to the $l$-th OAM-mode can be expressed as follows:
     \begin{equation}
        \gamma_{l}\!=\! \!\frac{\mathbb{E}[\bm{s}_{l}\bm{s}_{l}^{H}]\bm{H}_{l}^{H}\bm{H}_{l}}{\mathbb{E}\!\left[\bm{w}_{l}\bm{w}_{l}^{H}
       \!\! +\!\!\rho^{2} \bm{\Omega}_{l} \bm{s}_{l}\bm{s}_{l}^{H}\bm{\Omega}_{l}^{H}
       \! \!+\!\! \rho^{2}\bm{\Omega}_{l}\bm{H}_{l}^{-1}\bm{w}_{l}\bm{w}_{l}^{H}(\!\bm{H}_{l}^{-1}\!)^{H}\!\bm{\Omega}_{l}^{H}
        \!\right]}.
     \end{equation}
     The loss, denoted by $\Delta_{l} $, between the perfect channel estimation and the imperfect channel estimation, for the $l$-th OAM-mode can be calculated as Eq.~\eqref{eq:loss}.
     Clearly, with the increase of $\rho$, the channel SNR loss increases as shown in Eq.~\eqref{eq:loss}. Therefore, the capacity under perfect channel estimation is higher than that under imperfect channel estimation.


The maximum capacity, denoted by $C$, using the conventional water-filling power allocation scheme where the better channel is allocated more power and the worse channel is allocated less power is calculated as follows:
\setcounter{equation}{51}
\begin{equation}
    C=\mathbb{E}_{\gamma}\left\{\sum_{m=0}^{M-1}\sum_{l=\lfloor\frac{-N+2}{2}\rfloor}^{\lfloor\frac{N}{2}\rfloor}\!\log_{2}\!\left[\!1\!+\!\frac{h_{l,m}^{2}}{\sigma_{l,m}^{2}}\left(\frac{1}{\mu^{*}}-\frac{\sigma_{l,m}^{2}}
    {h_{l,m}^{2}}\right)^{+}\right]\right\},
    \label{eq:C_max}
\end{equation}
where $\mathbb{E}_{\gamma}(\cdot)$ denotes the expectation regarding instantaneous received SNR, $\sigma_{l,m}^{2}$ is the variance of noise for the $(l,m)$-th OAM-subcarrier block, $\mu^{*}$ represents the optimal Lagrangian multiplier, and $(\cdot)^{+}= \max\{\cdot, 0\}$.

To compare the capacity of HODM communications with that of conventional OFDM communications, we express the maximum capacity, denoted by $C_{OFDM}$, of the conventional OFDM communications with the water-filling power allocation scheme as follows:
    \begin{eqnarray}
        C_{OFDM}=\mathbb{E}_{\gamma}\left\{\sum_{m=0}^{M-1}\log_{2}\left[1+\frac{h_{m}^{2}}{\sigma_{m}^{2}}\left(
        \frac{1}{\xi^{*}}-\frac{\sigma_{m}^{2}}{h_{m}^{2}}\right)^{+}\right]\right\},\label{eq:OFDM}
  \end{eqnarray}
where $h_{m}$ and $\sigma_{m}^{2}$ denote the channel amplitude gain and variance of the received noise, respectively, for the $m$-th subcarrier. Also, $\xi^{*}$ represents the optimal Lagrangian multiplier.

Compared with Eqs. \eqref{eq:C_max} and \eqref{eq:OFDM}, HODM communications achieve both OAM multiplexing and frequency multiplexing while OFDM communications only achieve frequency multiplexing. Thus, the capacity of the HODM scheme is higher than that of the OFDM scheme in wireless communications.

\section{Performance Analysis}\label{sec:performance}

In this section, numerical simulation results are presented to evaluate the performance of wireless communication with our developed HODM scheme in sparse multipath environments. Firstly, Section~\ref{subsec:gain} evaluates the channel amplitude gain of our developed HODM scheme. Secondly, Section~\ref{subsec:before} depicts the optimal allocated power and corresponding capacities before converging. Throughout the evaluations, we set the all permittivities as 15, the first subcarrier frequency as 60 GHz, the bandwidth of each subcarrier as 5 MHz, the Rician shape parameter as $10$ dB, and the average total transmit power as 2 W. In Sections~\ref{subsec:before}, the distance $D$ is set as 3 m.


\subsection{Channel Amplitude Gain}\label{subsec:gain}

\begin{figure}
  \centering
  \includegraphics[width=0.52\textwidth]{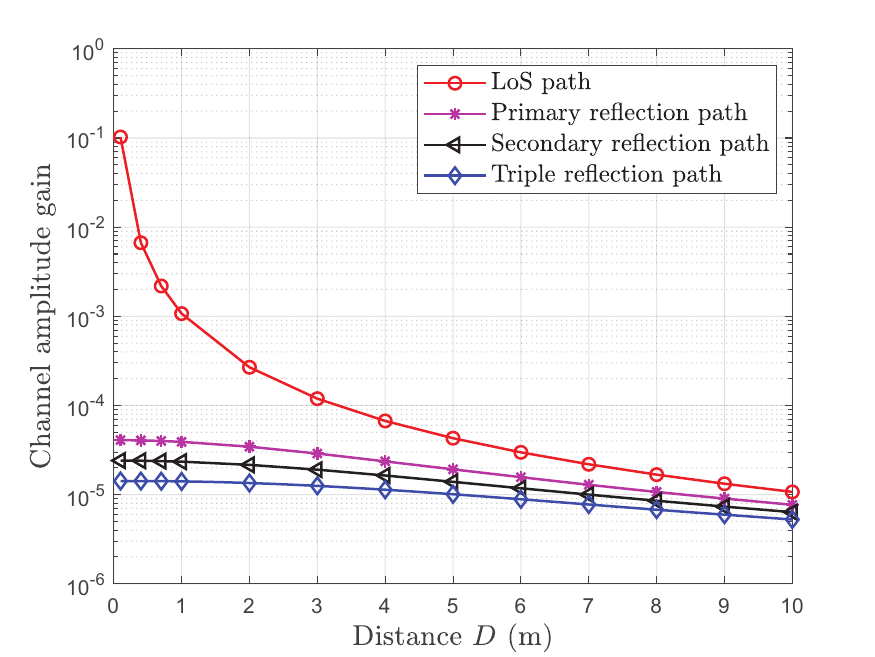}\\
  \caption{The channel amplitude gains of different paths for OAM-mode 1.}
  \label{fig:gain_path}
    \vspace{-10pt}
\end{figure}

Figure~\ref{fig:gain_path} shows the channel amplitude gains of different paths for OAM-mode 1, where we set a primary reflection path, a secondary reflection path, a reflection path, and $N=M=8$. Observing Fig.~\ref{fig:gain_path}, the channel amplitude gain decreases as the distance $D$ increases for all paths. This proves that energy of waves declines with the increase of distance. Also, the channel amplitude gain of reflection paths is much smaller than that of the LoS path. This result is caused by the specular reflectors, thus leading to energy attenuation. In addition, the channel amplitude gain of the LoS path drops rapidly within a certain distance, which is due to the first kind Bessel function.

In the sparse multipath environments, the total transmission paths, denoted by $L_{p}$, is given as follows:
\begin{equation}
    L_{p}=1+N_{r}+N_{t}+N_{e}.
\end{equation}
\begin{figure}
  \centering
  \includegraphics[width=0.52\textwidth]{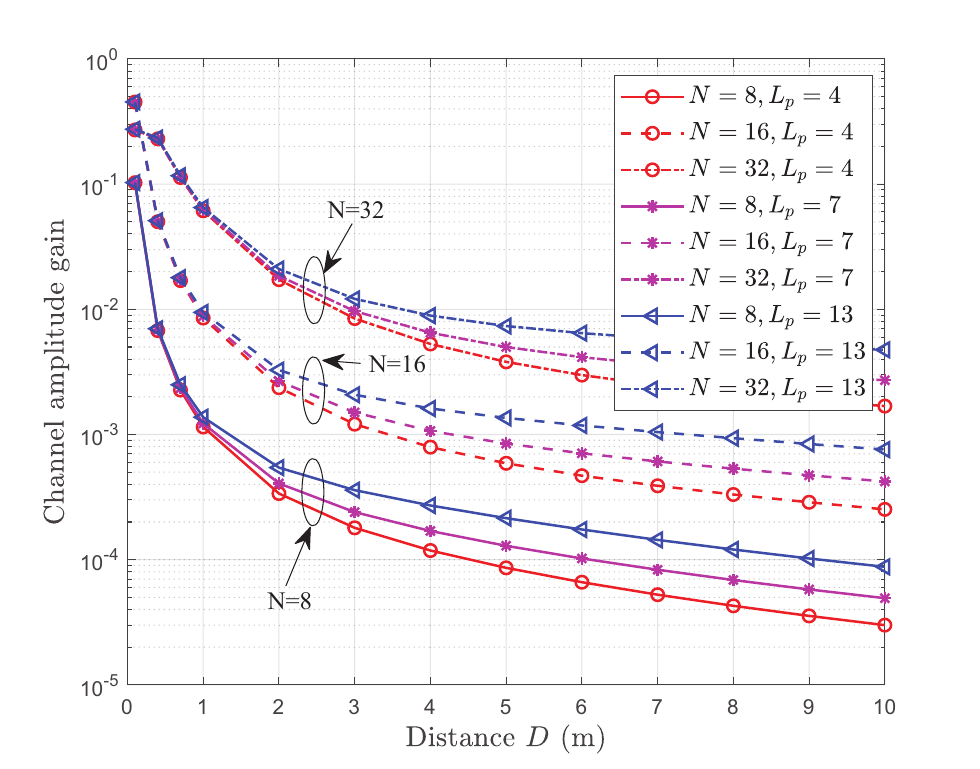}\\
  \caption{The channel amplitude gain of OAM-mode 1 versus different number of OAM-modes and transmission paths.}
  \label{fig:gain_N}
    \vspace{-10pt}
\end{figure}
\begin{figure}
  \centering
  \includegraphics[width=0.53\textwidth]{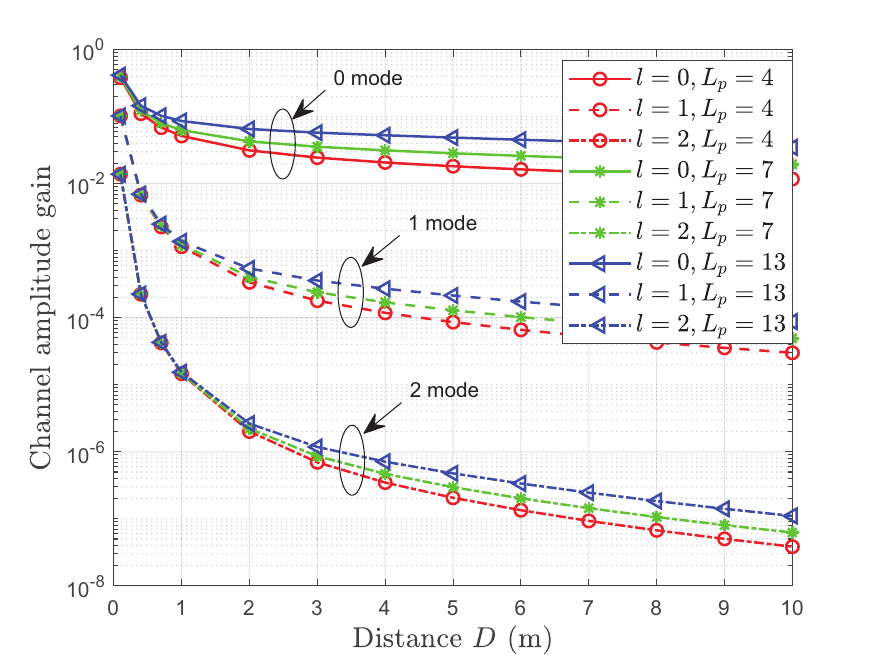}\\
  \caption{The channel amplitude gains of different order OAM-mode.}
  \label{fig:gain_L}
    \vspace{-10pt}
\end{figure}
Figure~\ref{fig:gain_N} depicts the channel amplitude gain of OAM-mode 1 versus different number of OAM-modes and transmission paths, where $L_{p}=$ 4, 7, and 13, respectively. Also, the number of narrowbands is selected as 8 and the number of OAM-modes is selected as 8, 16, as well as 32, respectively. Clearly, with  the total number of OAM-modes increasing, the channel amplitude gain monotonically goes up. Channel amplitude gains are propitiation to the total number of OAM-modes, which can be proved by Eqs.~\eqref{eq:y_ml_los}, \eqref{eq:y_ml_pri}, \eqref{eq:y_ml_sec}, and \eqref{eq:y_ml_thi}. Also, with the number of transmission paths increasing, channel amplitude gains of reflection paths increase, thus resulting in increasing the channel amplitude gains in wireless communications.

Figure~\ref{fig:gain_L} depicts the channel amplitude gains of different order OAM-modes, where the total number of OAM-modes and narrowbands is set as 8, respectively. Also, the number of paths is set as 4, 7, and 13, respectively. Obviously, the channel amplitude gain is higher with low order OAM-mode than that with high order OAM-mode as depicts in Fig.~\ref{fig:gain_L}. Because the radiuses of transmit UCA and receive UCA is far smaller than the distance $D$, the variable in Bessel function is very small. Thus, the value of Bessel function is smaller with higher order OAM-modes. Hence, this result proves that the OAM-based waves divergent faster with high order OAM-modes than that with low order OAM-modes. Moreover, the channel amplitude gain decreases as transmission distance increases.

\subsection{Power Allocation and Capacities Before Converging}\label{subsec:before}

\begin{figure}
  \centering
  \includegraphics[width=0.52\textwidth]{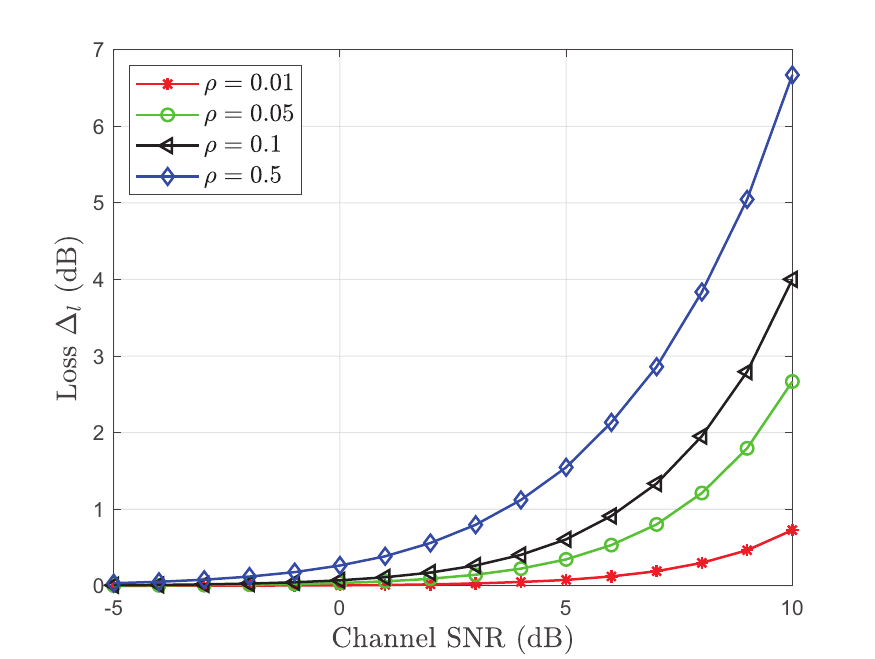}\\
  \caption{The SNR loss versus SNR.}
  \label{fig:loss}
    \vspace{-10pt}
\end{figure}

To analyze the impact of CEE on the performance of radio vortex wireless communications, we present the channel SNR loss versus the channel SNR in Fig.~\ref{fig:loss}, where we set the accuracy of channel estimation $\rho$ as 0.01, 0.05, 0.1, and 0.5, respectively. As shown in Fig.~\ref{fig:loss}, the SNR loss increases as $\rho$ increases, which can be proved by Eq.~\eqref{eq:loss}. For example, the loss is 6.7 dB and 0.7 dB with respect to $\rho=0.5$ and $\rho=0.01$, respectively, at 10 dB channel SNR. Also, the loss increases as the channel SNR increases. Thus, extra power is needed to reach the expected channel SNR under imperfect channel estimation.

\begin{figure}
  \centering
  \includegraphics[width=0.52\textwidth]{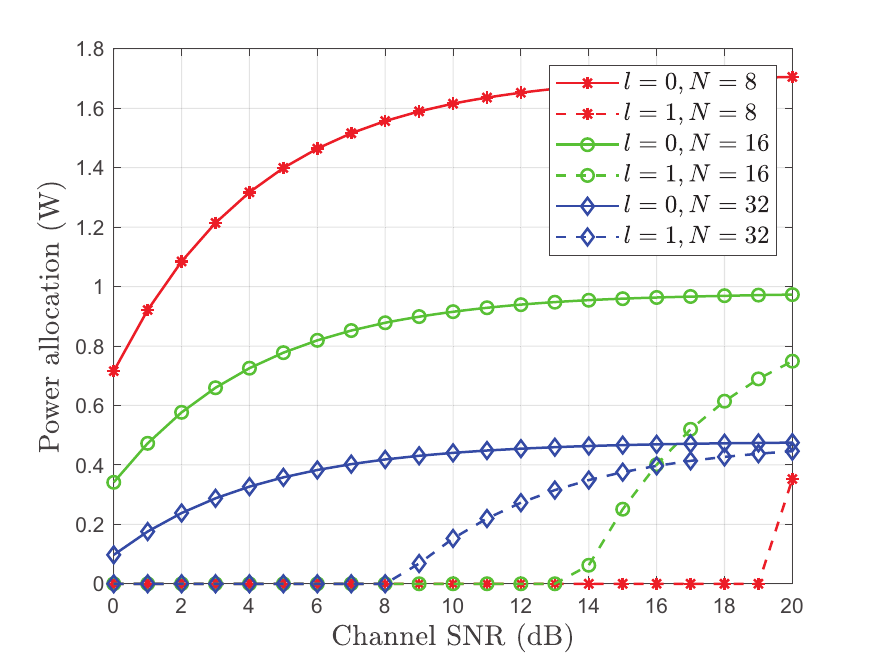}
  \caption{The optimal power allocation schemes with different number of OAM-modes before converging.}
  \label{fig:power_N}
    \vspace{-10pt}
\end{figure}

In Fig.~\ref{fig:power_N}, the optimal power allocation schemes are plotted with different number of OAM-modes versus channel SNR before converging, where the number of paths is set as 4. Also, we set $N = M =$ 8, 16, and 32, respectively. Observing Eq.~\eqref{eq:C_max}, we can see that the maximum capacity is obtained after the expectation operation regarding instantaneous received SNR. For any given channel state information, the sum of allocated instantaneous power over OAM-modes/subcarriers is equal to the total transmit power at the transmitter, but not to 2 watts. Thus, we can obtain the maximum instantaneous capacity of our proposed HODM scheme. Since a LoS path coexists with several reflection paths, the channels follow the Rician distribution with the average power 2 W constraint. If the sum of allocated power is always equal to 2 W in the whole SNR region, the obtained average capacity is not maximum. To maximize the average capacity of our proposed HODM scheme in sparse multipath environments, the sum of allocated power over modes/subcarriers at the transmitter is a Rician random variable. Since channel amplitude gains with respect to high order OAM-mode is much smaller, they are allocated to fewer power. Therefore, Fig.~\ref{fig:power_N} only shows the allocated power of OAM-modes 0 and 1. Clearly, the increase of channel SNR makes the increase of allocated power. Moreover, the channel with OAM-mode 0 is allocated to higher power than that with OAM-mode 1, because the channel amplitude gain of OAM-mode 0 is much larger than that of OAM-mode 1. Furthermore, since the channel amplitude gains of both OAM-modes 0 and 1 increase with the total number of OAM-modes increasing, the optimal allocated power of OAM-mode 0 decreases in the whole SNR region and the allocated power of OAM-mode 1 increases in the low SNR region. As $N$ increases, at a given high SNR the allocated power of OAM-mode 1 first increases because the almost all the power is allocated to low order OAM-modes and then decreases because a part of the power is allocated to other higher order OAM-modes. 

\begin{figure}
  \centering
  \includegraphics[width=0.52\textwidth]{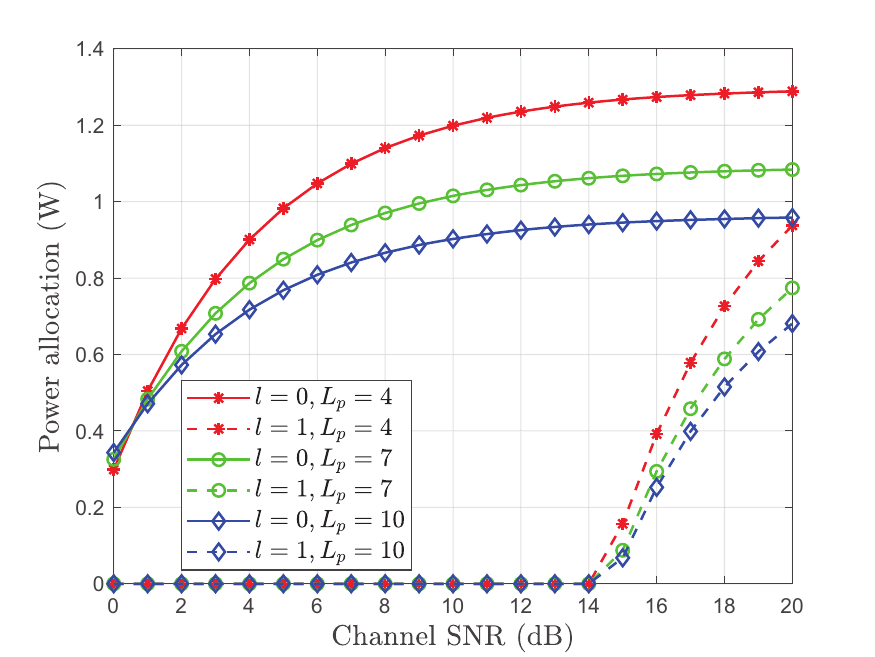}\\
  \caption{The optimal power allocation schemes with different number of paths before converging.}
  \label{fig:power_L}
    \vspace{-10pt}
\end{figure}

Figure~\ref{fig:power_L} presents the optimal power allocation schemes with different number of paths before converging, where the total number of OAM-modes and narrowbands is set as 16, respectively. In addition, the number of transmission paths is set as 4, 7, and 10, respectively. 
Observing Fig.~\ref{fig:power_L}, the increase of transmission paths leads to the reduction of optimal allcoated power. The reason is given as that the channel amplitude gain of all OAM-modes increases as the number of transmission paths rises. For few paths, the low order OAM-modes are allocated to most power while the higher order OAM-modes can be allocated more power with more paths. Hence, the high order OAM-modes can play important roles in the capacity with a large number of transmission paths before converging in wireless communications.

\begin{figure}
  \centering
  \includegraphics[width=0.52\textwidth]{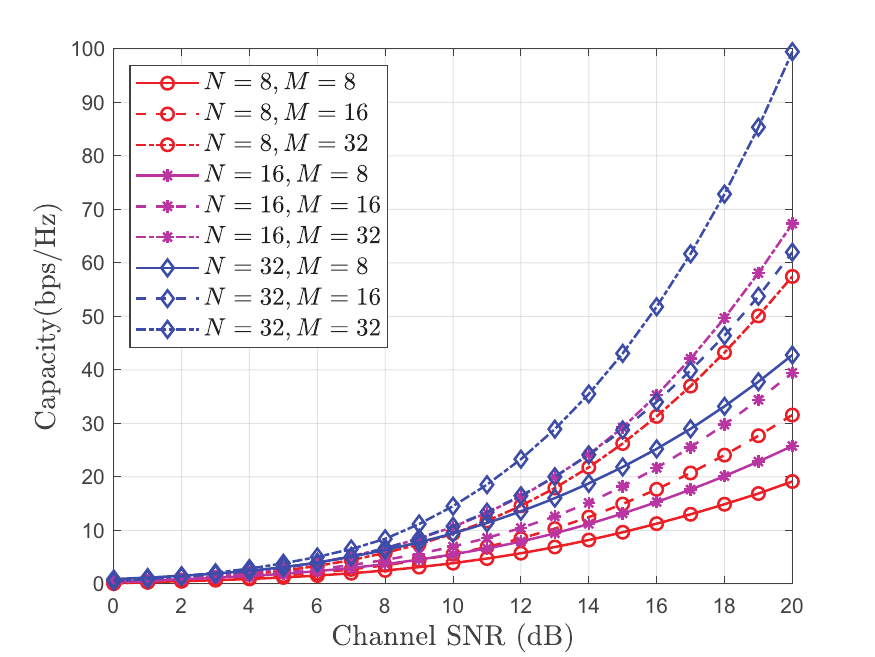}\\
  \caption{The capacities with different number of narrowbands and OAM-modes, respectively, in HODM communications.}
  \label{fig:capacity_N}
    \vspace{-10pt}
\end{figure}

Figure~\ref{fig:capacity_N} shows the maximum capacities by using water-filling shceme with different number of narrowbands and OAM-modes in HODM communications, where $L_{p}=4$. Also, the number of narrowbands and OAM-modes is equal to 8, 16, and 32, respectively. Clearly, the achievable maximum capacities by using our developed scheme monotonically goes up with the growing channel SNR as illustrated in Fig.~\ref{fig:capacity_N}. This is consistent with the optimal power allocation as shown in Fig.~\ref{fig:power_N}. Moreover, the increase of OAM-modes and narrowbands induces the benifit to the total number of independent and parallel transmission channels, thus making the increase of maximum capacities in HODM communications. Furthermore, we can find that the maximum capacities cannot proportionally increase with the OAM-modes increasing due to the divergence of high order OAM-waves in radio vortex wireless communications. Thereby, OAM-waves convergence to significantly increase the capacities is a challenge in future.

\begin{figure}
  \centering
  \includegraphics[width=0.52\textwidth]{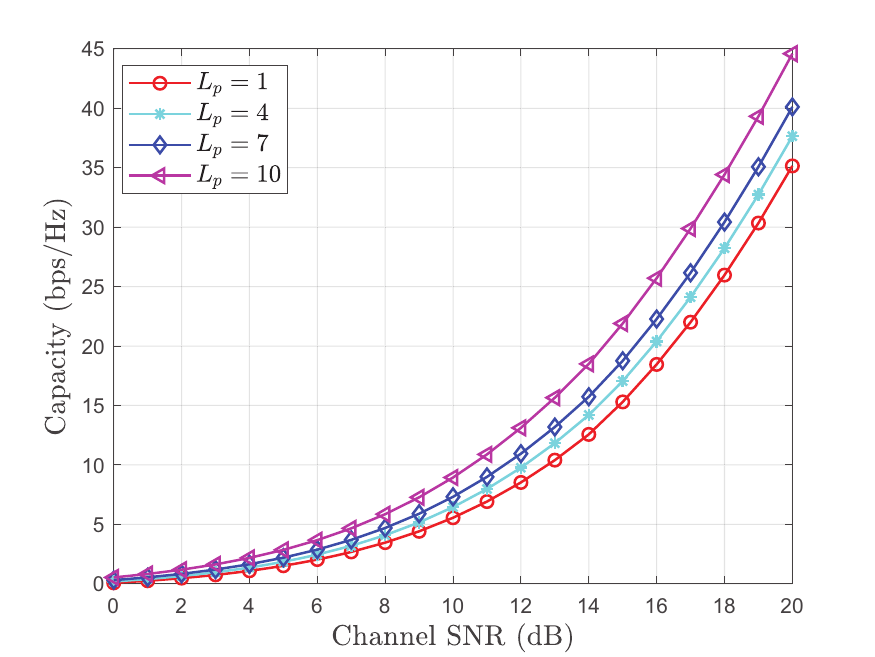}\\
 \caption{The comparisons of capacities before converging with different number of paths.}
  \label{fig:capacity_L}
    \vspace{-10pt}
\end{figure}
%

In Fig.~\ref{fig:capacity_L}, we present the maximum capacities by using conventional water-filling scheme in HODM scheme with different paths, where the total number of narrowbands and OAM-modes is 16, respectively. The number of transmission paths is set as 1, 4, 7, and 10, respectively. One path implies that there is a LoS path for signal transmission. Fig.~\ref{fig:capacity_L} illustrates that the maximum capacity goes up as the total number of transmission paths increases when the number of narrowbands and OAM-modes is given. This is because the channel amplitude gains of high order OAM-modes increase, thus resulting in more power assigned to these high order OAM-modes and less power assigned to low order OAM-modes.

\begin{figure}
  \centering
  \includegraphics[width=0.52\textwidth]{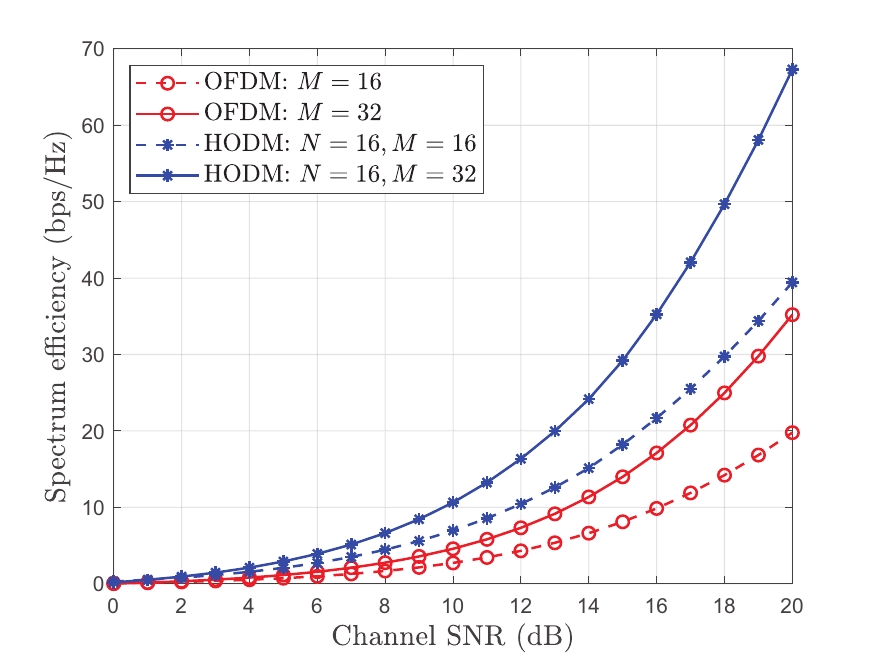}\\
  \caption{Comparison of capacity between our proposed HODM scheme and the conventional OFDM scheme.}
  \label{fig:capacity_com}
    \vspace{-10pt}
\end{figure}

\begin{figure*}[hb]
\hrulefill
\setcounter{equation}{54}
\begin{eqnarray}
\begin{aligned}
    &\frac{1}{MN}\sum_{k=0}^{M-1} \sum_{v=0}^{N-1} \sum_{n=0}^{N-1} h_{vn,LoS,k}X_{n,k} e^{-j \frac{2\pi vl}{N}}e^{-j\frac{2\pi mk}{M}}
\\
    &=\frac{1}{MN}\sum_{k=0}^{M-1} \sum_{v=0}^{N-1}\left(\sum_{n=0}^{N-1} h_{vn,LoS,k}\sum_{m^{'}=0}^{M-1}\sum_{l^{\prime}=\lfloor\frac{-N+2}{2}\rfloor}^{\lfloor\frac{N}{2}\rfloor} s_{l^{'},m^{'}} e^{j\frac{2\pi nl^{'}}{N}} e^{j\frac{2\pi km^{'}}{M}}
    \right) e^{-j\frac{2\pi vl}{N}}e^{-j\frac{2\pi km}{M}}
\\
    & \overset{N \rightarrow \infty}{\approx} \sum_{m^{'}=0}^{M-1}\sum_{l^{\prime}=\lfloor\frac{-N+2}{2}\rfloor}^{\lfloor\frac{N}{2}\rfloor} s_{l^{'},m^{'}} \frac{1}{MN} \sum_{k=0}^{M-1} \sum_{v=0}^{N-1} \frac{\beta \lambda e^{-j\frac{2\pi \sqrt{D^{2}+r_{1}^{2}+r_{2}^{2}}}{\lambda}}}{4\pi\sqrt{D^{2}+r_{1}^{2}+r_{2}^{2}}}  \frac{N}{2\pi} \int_{0}^{2\pi} e^{j\frac{2\pi r_{1}r_{2}\cos{\phi^{'}}}{\lambda \sqrt{D^{2}+r_{1}^{2}+r_{2}^{2}}}} e^{-j\phi^{'} l^{'}} d \phi^{'} e^{j\frac{2\pi v (l^{'}-l)}{N}} e^{j\frac{2\pi k (m^{'}-m)}{M}}
\\
     &= \sum_{m^{'}=0}^{M-1}\sum_{l^{\prime}=\lfloor\frac{-N+2}{2}\rfloor}^{\lfloor\frac{N}{2}\rfloor} s_{l^{'},m^{'}} \frac{1}{MN} \sum_{k=0}^{M-1} \sum_{v=0}^{N-1}  \frac{\beta \lambda e^{-j\frac{2\pi \sqrt{D^{2}+r_{1}^{2}+r_{2}^{2}}}{\lambda}}}{4\pi\sqrt{D^{2}+r_{1}^{2}+r_{2}^{2}}} N j^{-l^{'}} J_{l^{'}}\left(\frac{2\pi r_{1} r_{2}}{\lambda \sqrt{D^{2}+r_{1}^{2}+r_{2}^{2}}}\right) e^{j\frac{2\pi v (l^{'}-l)}{N}} e^{j\frac{2\pi k (m^{'}-m)}{M}}
\\
&=\left\{\begin{array}{lcl}
     \underbrace{\frac{\beta \lambda_{m} N j^{-l} e^{-j\frac{2\pi \sqrt{D^{2}+r_{1}^{2}+r_{2}^{2}}}{\lambda_{m}}}}{4\pi\sqrt{D^{2}+r_{1}^{2}+r_{2}^{2}}}  J_{l}\left(\frac{2\pi r_{1} r_{2}}{\lambda_{m} \sqrt{D^{2}+r_{1}^{2}+r_{2}^{2}}}\right)}_{h_{l,m,LoS}} s_{l,m} & & {m=m^{'}\ {\rm{and}} \ l=l^{'};}\\
0 & & {\rm{otherwise.}}
\end{array} \right.
\end{aligned}
\label{eq:y_ml_los}
\end{eqnarray}
\end{figure*}

Figure~\ref{fig:capacity_com} compares the capacities between our proposed HODM scheme and the conventional OFDM scheme in wireless communications versus the channel SNR, where the available subcarriers $M$ is 16 and 32, respectively, in OFDM and HODM communications. 
Also, we set the available OAM-modes as $N=16$. As shown in Fig.~\ref{fig:capacity_com}, the capacities of both conventional OFDM scheme and our developed HODM scheme increase as the channel SNR increases. Clearly, the capacity of our developed HODM scheme is higher than that of the conventional OFDM scheme when the number of subcarriers of the HODM scheme is same with that of the OFDM scheme. 
This is because the OAM-modes bring the increase of capacity in wireless communications. These results verify that our developed HODM can be used for higher capacity of wireless communications in sparse multipath environments.

\section{Conclusions} \label{sec:conc}

In this paper, a joint OAM multiplexing and OFDM scheme called HODM is proposed to achieve high capacity while resisting multipath interference for wireless communications in sparse multipath environments. Firstly, we built the OAM-based wireless channel model for sparse multipath transmission comprising a LoS path and several reflection paths, where the four-time or more reflection signals are ignored. Secondly, we proposed to compensate the phase difference caused by the different length of paths to mitigate the inter-mode interference in HODM communications. Then, we introduced the conventional water-filling algorithm to maximize capacity of our developed HODM scheme in radio vortex wireless communications. Numerical and theorical results have been provided to verify the significant increase of capacity in sparse multipath environments by using our developed HODM scheme.

\begin{appendices}
\section{Proof of Theorem 1}\label{Appendix:_A}
The first term on the right hand of Eq.~\eqref{eq:y_ml_exp} corresponding to the LoS path can be derived as Eq.~\eqref{eq:y_ml_los}. Thus, the channel amplitude gain of LoS path with respect to the $(l,m)$-th OAM-subcarrier block can be obtained.

\begin{figure*}
\setcounter{equation}{56}
\begin{eqnarray}
\begin{aligned}
    &\frac{1}{MN}\sum_{k=0}^{M-1} \sum_{v=0}^{N-1}\sum_{nr=1}^{N_{r}} \sum_{n=0}^{N-1} h_{vn,nr,k}h_{vn,e}\widetilde{X}_{n-(L_{r}-nr),k-(L_{r}-nr)}^{'} e^{-j \frac{2\pi vl}{N}}e^{-j\frac{2\pi mk}{M}}
\\
& \approx \sum_{m^{'}=0}^{M-1}\sum_{l^{\prime}=\lfloor\frac{-N+2}{2}\rfloor}^{\lfloor\frac{N}{2}\rfloor} s_{l^{'},m^{'}} \frac{1}{MN} \sum_{k=0}^{M-1} \sum_{v=0}^{N-1} \sum_{nr=1}^{N_{r}} \!\!\frac{ R_{nr}\beta \lambda e^{-j\frac{2\pi \sqrt{D^{2}+r_{1}^{2}+r_{2}^{2}+4d_{nr}^{2}}}{\lambda}}}{4\pi\sqrt{D^{2}+r_{1}^{2}+r_{2}^{2}+4d_{nr}^{2}}}  e^{-j\frac{2\pi r_{1}r_{2}\cos\left[\left(\frac{2\pi n}{N}\right)+\left(\frac{2\pi v}{N}\right)\right]}{\lambda \sqrt{D^{2}+r_{1}^{2}+r_{2}^{2}+4d_{nr}^{2}}}}
e^{-j \frac{2\pi l^{\prime}[n-(L_{r}-n_{r})]}{N}}
\\
&\hspace{11cm}e^{j\frac{2\pi m^{\prime}[k-(L_{r}-n_{r})]}{M}} e^{-j \frac{2\pi vl}{N}}e^{-j\frac{2\pi mk}{M}}
\\
 & \overset{(a)}{\approx} \sum_{m^{'}=0}^{M-1}\sum_{l^{\prime}=\lfloor\frac{-N+2}{2}\rfloor}^{\lfloor\frac{N}{2}\rfloor} s_{l^{'},m^{'}} \frac{1}{MN} \sum_{k=0}^{M-1} \sum_{v=0}^{N-1} \sum_{nr=1}^{N_{r}} \frac{ R_{nr}\beta \lambda e^{-j\frac{2\pi \sqrt{D^{2}+r_{1}^{2}+r_{2}^{2}+4d_{nr}^{2}}}{\lambda}}}{4\pi\sqrt{D^{2}+r_{1}^{2}+r_{2}^{2}+4d_{nr}^{2}}}  \frac{N}{2\pi} \int_{0}^{2\pi} e^{j\frac{2\pi r_{1}r_{2}\cos{\phi^{'}}}{\lambda \sqrt{D^{2}+r_{1}^{2}+r_{2}^{2}+4d_{nr}^{2}}}} e^{-j\phi^{'} l^{'}} d \phi^{'}
 \\
&\hspace{9cm}(-1)^{l^{\prime}}e^{j \frac{2\pi l^{\prime}(L_{r}-n_{r})}{N}}e^{-j\frac{2\pi m^{\prime}(L_{r}-n_{r})}{M}} e^{j \frac{2\pi v(l^{\prime}-l)}{N}}e^{j\frac{2\pi (m^{\prime}-m)k}{M}}
\\
&=\left\{\begin{array}{lcl}
  \underbrace{\sum\limits_{nr=1}^{N_{r}} \frac{ R_{nr}\beta \lambda_{m} N j^{-3l} e^{-j\frac{2\pi \sqrt{D^{2}+r_{1}^{2}+r_{2}^{2}+4d_{nr}^{2}}}{\lambda_{m}}}}{4\pi\sqrt{D^{2}+r_{1}^{2}+r_{2}^{2}+4d_{nr}^{2}}}  J_{l}\left(\frac{2\pi r_{1} r_{2}}{\lambda_{m} \sqrt{D^{2}+r_{1}^{2}+r_{2}^{2}+4d_{nr}^{2}}}\right) e^{j\frac{2\pi l (L_{r}-nr)}{N}} e^{-j\frac{2\pi m (L_{r}-nr)}{M}}}_{h_{l,m,N_{r}}} s_{l,m}
\\
     \hspace{13cm}  {m=m^{'}\ {\rm{and}} \ l=l^{'};}
\\
   0  \hspace{12.8cm} {\rm{otherwise,}}
\end{array} \right.
\end{aligned}
\label{eq:y_ml_pri}
\end{eqnarray}
\hrulefill
  \vspace{-10pt}
\end{figure*}

\section{Proof of Theorem 2}\label{Appendix:_C}

We assume that $q^{2}=D^{2}+r_{1}^{2}+r_{2}^{2}$. According to $d_{nr}, d_{max} \ll D$ and phase compensation factor, with respect to the $nr$-th primary reflected path from the $n$-th array to the $v$-th receive array transmission, we have
\setcounter{equation}{55}
\begin{eqnarray}
  & & h_{vn,e} e^{j\frac{2\pi \left[2r_{1}h_{nr}\cos\left(\frac{2\pi n}{N}\right)+2r_{2}d_{nr}\cos\left(\frac{2\pi v}{N}\right)\right]}{\lambda \sqrt{D^{2}+r_{1}^{2}+r_{2}^{2}+4d_{nr}^{2}}}}
\nonumber\\
  && = e^{j\frac{2\pi}{\lambda} \left[r_{2} \cos\left(\frac{2\pi v}{N}\right)+r_{1}\cos\left(\frac{2\pi n}{N}\right)\right]\left(\frac{2d_{nr}}{\sqrt{q^{2}+ 4d_{nr}^{2}}}-\frac{2d_{max}}{\sqrt{q^{2}+4d_{max}^{2}}}\right)}
\nonumber\\
  & &\approx 1.
\end{eqnarray}
Thus, with the method of 2D-FFT algorithm, the demodulated signal of the second right hand of Eq.~\eqref{eq:y_vk_com} is expressed as Eq.~\eqref{eq:y_ml_pri}, where $(a)$ represents $2\pi n/N +2\pi v/N = \phi^{'} + \pi$ and $N \rightarrow \infty$. Similar to the analysis of Eq.~\eqref{eq:y_ml_pri}, we can also obtain the corresponding demodulated signals of the secondary reflection paths and triple reflection paths as shown in Eqs.~\eqref{eq:y_ml_sec} and \eqref{eq:y_ml_thi}, respectively. Then, we can obtain the channel amplitude gain of reflection paths for the $(l,m)$-th OAM-subcarrier block.

\begin{figure*}
\setcounter{equation}{57}
\begin{eqnarray}
\begin{aligned}
    &\frac{1}{MN}\sum_{k=0}^{M-1} \sum_{v=0}^{N-1}\sum_{nt=1}^{N_{t}} \sum_{n=0}^{N-1} h_{vn,nt,k} h_{vn,e} X_{n-(L_{t}-nt),k-(L_{t}-nt)} e^{-j \frac{2\pi vl}{N}}e^{-j\frac{2\pi mk}{M}}
\\
&\approx\left\{\begin{array}{lcl}
    \underbrace{\sum\limits_{nt=1}^{N_{t}}\frac{ R_{nt}\beta \lambda_{m} N j^{-l} e^{-j\frac{2\pi \sqrt{D^{2}+r_{1}^{2}+r_{2}^{2}+4d_{nt}^{2}}}{\lambda_{m}}}}{4\pi\sqrt{D^{2}+r_{1}^{2}+r_{2}^{2}+4d_{nt}^{2}}}  J_{l}\left(\frac{2\pi r_{1} r_{2}}{\lambda_{m} \sqrt{D^{2}+r_{1}^{2}+r_{2}^{2}+4d_{nt}^{2}}}\right) e^{-j\frac{2\pi l (L_{t}-nt)}{N}} e^{-j\frac{2\pi m (L_{t}-nt)}{M}}}_{h_{l,m,N_{t}}} s_{l,m}
\\
     \hspace{13cm}  {m=m^{'}\ {\rm{and}} \ l=l^{'};}
\\
   0  \hspace{12.8cm} {\rm{otherwise.}}
\end{array} \right.
\end{aligned}
\label{eq:y_ml_sec}
\end{eqnarray}
\hrulefill
  \vspace{-10pt}
\end{figure*}

\begin{figure*}
\setcounter{equation}{58}
\begin{equation}
\begin{aligned}
    &\frac{1}{MN}\sum_{k=0}^{M-1} \sum_{v=0}^{N-1}\sum_{ne=1}^{N_{e}} \sum_{n=0}^{N-1} h_{vn,ne,k}h_{vn,e} \widetilde{X}_{n-(L_{e}-ne),k-(L_{e}-ne)}^{'} e^{-j \frac{2\pi vl}{N}}e^{-j\frac{2\pi mk}{M}}
\\
&\approx\left\{\begin{array}{lcl}
    \underbrace{\sum\limits_{ne=1}^{N_{e}} \frac{ R_{ne}\beta \lambda_{m} N j^{-3l} e^{-j\frac{2\pi \sqrt{D^{2}+r_{1}^{2}+r_{2}^{2}+4d_{ne}^{2}}}{\lambda_{m}}}}{4\pi\sqrt{D^{2}+r_{1}^{2}+r_{2}^{2}+4d_{ne}^{2}}}  J_{l}\left(\frac{2\pi r_{1} r_{2}}{\lambda_{m} \sqrt{D^{2}+r_{1}^{2}+r_{2}^{2}+4d_{ne}^{2}}}\right) e^{j\frac{2\pi l (L_{e}-ne)}{N}} e^{-j\frac{2\pi m (L_{e}-ne)}{M}}}_{h_{l,m,N_{e}}} s_{l,m}
\\
     \hspace{13cm}  {m=m^{'}\ {\rm{and}} \ l=l^{'};}
\\
   0  \hspace{12.8cm} {\rm{otherwise.}}
\end{array} \right.
\end{aligned}
\label{eq:y_ml_thi}
\end{equation}
\hrulefill
  \vspace{-10pt}
\end{figure*}
\end{appendices}
\bibliographystyle{IEEEbib}
\bibliography{References}
\begin{IEEEbiography}[{\includegraphics[width=1in,height=1.25in,clip,keepaspectratio]{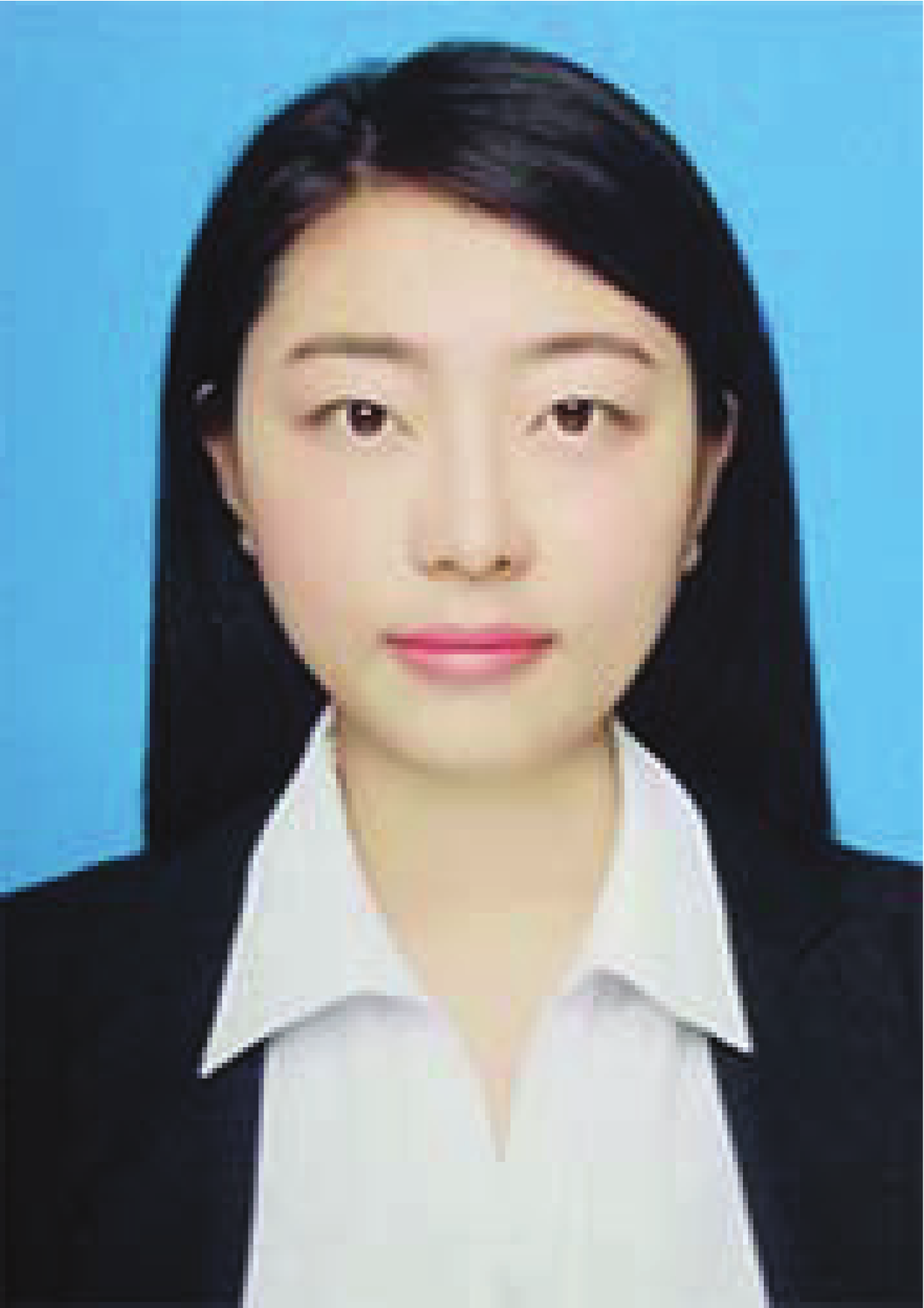}}]{Liping Liang} received B.S. degree in Electronic and Information Engineering from Jilin University, Changchun, China in 2015. She is currently working towards the Ph.D. degree in communication and information system from Xidian University, Xi'an, China. Her research interests focus on 5G wireless communications with emphasis on radio vortex wireless communications and anti-jamming communications.
\end{IEEEbiography}

\begin{IEEEbiography}[{\includegraphics[width=1in,height=1.25in,clip,keepaspectratio]{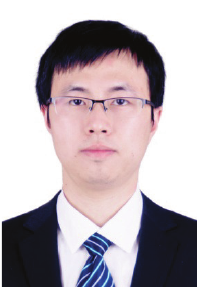}}]{Wenchi Cheng} (M'14-SM'18) received the B.S. and Ph.D. degrees in telecommunication engineering from Xidian University, Xian, China, in 2008 and 2013, respectively, where he is a Full Professor. He was a Visiting Scholar with Networking and Information Systems Laboratory, Department of Electrical and Computer Engineering, Texas A\&M University, College Station, TX, USA, from 2010 to 2011. His current research interests include B5G/6G wireless networks, emergency wireless communications, and orbital-angular-momentum based wireless communications. He has published more than 100 international journal and conference papers in IEEE Journal on Selected Areas in Communications, IEEE Magazines, IEEE Transactions, IEEE INFOCOM, GLOBECOM, and ICC, etc. He received URSI Young Scientist Award (2019), the Young Elite Scientist Award of CAST, the Best Dissertation of China Institute of Communications, the Best Paper Award for IEEE ICCC 2018, the Best Paper Award for IEEE WCSP 2019, and the Best Paper Nomination for IEEE GLOBECOM 2014. He has served or serving as the Associate Editor for IEEE Systems Journal, IEEE Communications Letters, IEEE Wireless Communications Letter, the IoT Session Chair for IEEE 5G Roadmap, the Wireless Communications Symposium Co-Chair for IEEE GLOBECOM 2020, the Publicity Chair for IEEE ICC 2019, the Next Generation Networks Symposium Chair for IEEE ICCC 2019, the Workshop Chair for IEEE ICC 2019/IEEE GLOBECOM 2019/INFOCOM 2020 Workshop on Intelligent Wireless Emergency Communications Networks.
\end{IEEEbiography}

\begin{IEEEbiography}[{\includegraphics[width=1in,height=1.25in,clip,keepaspectratio]{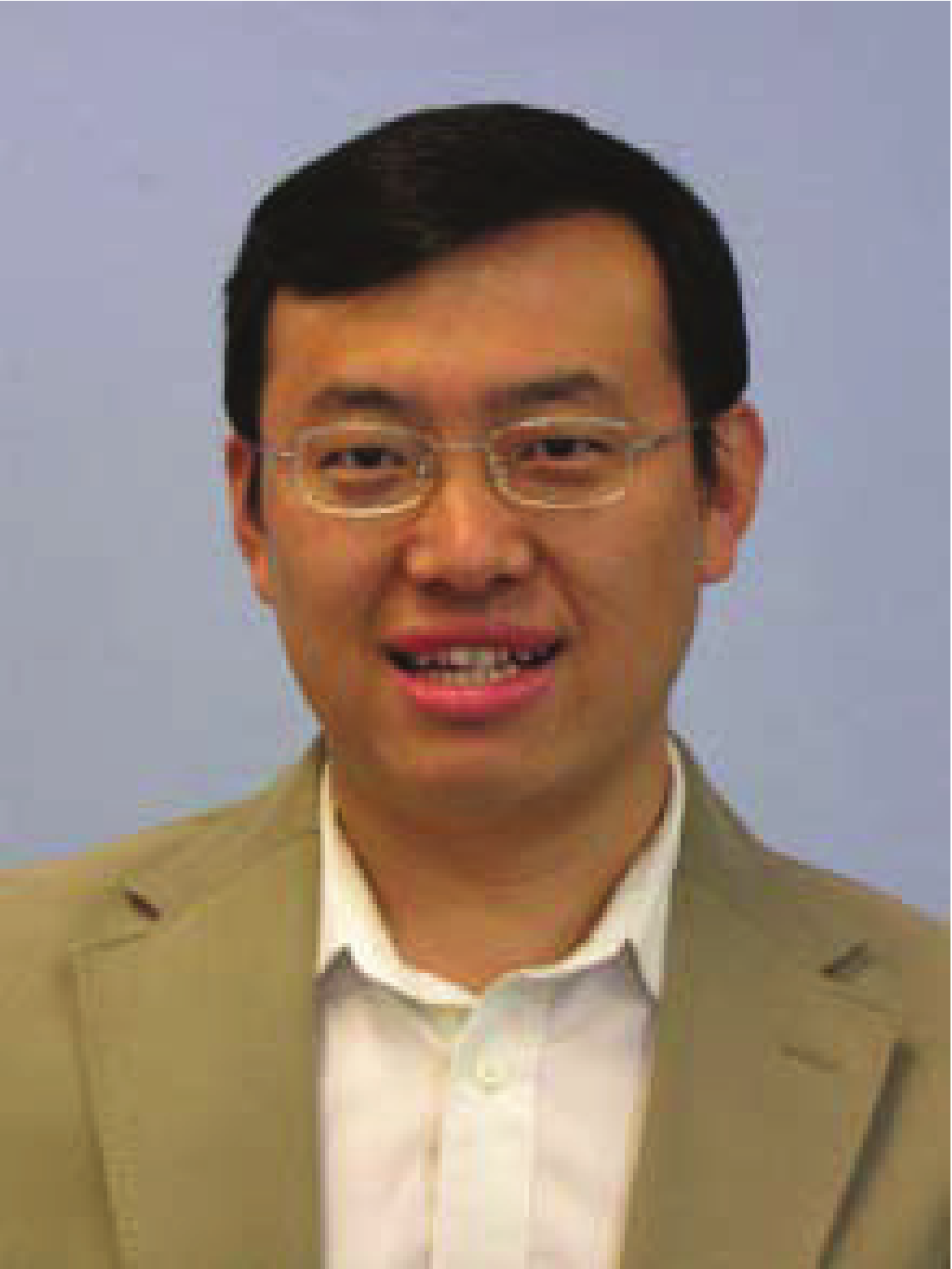}}]{Wei Zhang} (S'01-M'06-SM'11-F'15) received the Ph.D. degree in electronic engineering from Chinese University of Hong Kong, Hong Kong, in 2005. He was a Research Fellow with Hong Kong University of Science and Technology, Clear Water Bay, Hong Kong SAR, from 2006 to 2007. Currently, he is a Professor at School of Electrical Engineering and Telecommunications, University of New South Wales (UNSW), Sydney, Australia. His research interests include UAV communications, mmWave communications, space information networks, and massive MIMO. He is the Editor-in-Chief of Journal of Communications and Information Networks (JCIN). He also serves as Chair for IEEE Wireless Communications Technical Committee. He is a member of Board of Governors of IEEE Communications Society. He is a member of Fellow Evaluation Committee of IEEE Vehicular Technology Society.

\end{IEEEbiography}

\begin{IEEEbiography}[{\includegraphics[width=1in,height=1.25in,clip,keepaspectratio]{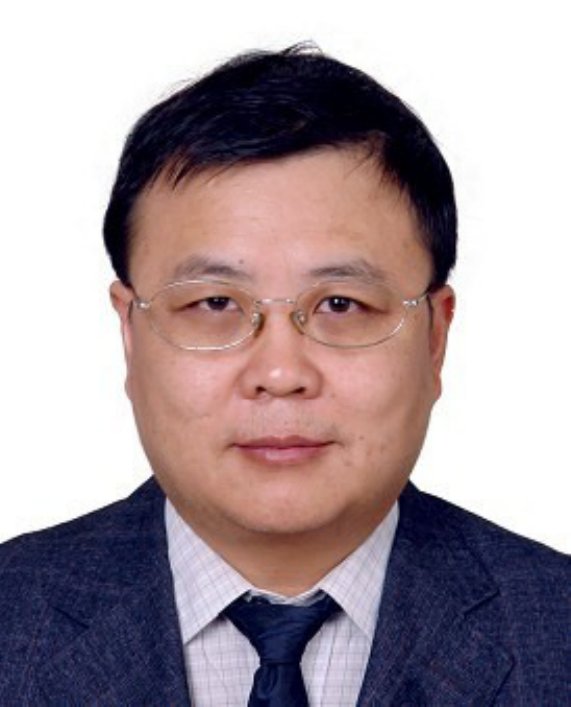}}]{Hailin Zhang} (M'97) received B.S. and M.S. degrees from Northwestern Polytechnic University, Xi'an, China, in 1985 and 1988 respectively, and the Ph.D. from Xidian University, Xi'an, China, in 1991. In 1991, he joined School of Telecommunications Engineering, Xidian University, where he is a senior Professor and the Dean of this school. He is also currently the Director of Key Laboratory in Wireless Communications Sponsored by China Ministry of Information Technology, a key member of State Key Laboratory of Integrated Services Networks, one of the state government specially compensated scientists and engineers, a field leader in Telecommunications and Information Systems in Xidian University, an Associate Director of National 111 Project. Dr. Zhang's current research interests include key transmission technologies and standards on broadband wireless communications for 5G and 5G-beyond wireless access systems. He has published more than 150 papers in journals and conferences.
\end{IEEEbiography}

\end{document}